\documentclass[12pt,aps,floatfix,nofootinbib]{revtex4}
\usepackage{amsmath}
\usepackage{latexsym}
\usepackage{float}
\usepackage{amssymb}
\usepackage{graphicx}
\usepackage{epsfig}
\newcommand{\di}{\mathrm{d}}

\newcommand{\DG}{\Delta G}

\newcommand{\al}{\alpha}
\newcommand{\be}{\beta}

\begin{document}

\title{
Controlling the temperature sensitivity of DNA--mediated colloidal
interactions through competing linkages
}

\author{
B.\ M.\ Mognetti$^1$, M.\ E.\ Leunissen$^{2}$, and D.\ Frenkel$^1$ \\
$^1$Department of Chemistry, University of Cambridge, Lensfield Road, Cambridge, CB2 1EW, United Kingdom \\
$^2$FOM Institute AMOLF, Science Park 104, 1098 XG, Amsterdam, The Netherlands
}

\begin{abstract}
We propose a new strategy to improve the self-assembly properties of DNA-functionalised colloids. The problem that we address is that DNA-functionalised colloids typically crystallize in a narrow temperature window, if at all. The underlying reason is the extreme sensitivity of DNA-mediated interactions to temperature or other physical control parameters. We propose to widen the window for colloidal crystallization by exploiting  the competition between DNA linkages with different nucleotide sequences, which results in a temperature-dependent switching of the dominant bond type. Following such a strategy, we can decrease the temperature dependence of DNA-mediated  self assembly to make systems that can crystallize in a wider temperature window than is possible with existing systems of DNA functionalised colloids. We report Monte Carlo simulations that show that the proposed strategy can indeed work in practice for real systems and specific, designable DNA sequences.  Depending on the length ratio of the different DNA constructs, we find that the bond switching is either energetically driven (equal length or `symmetric' DNA) or controlled by a combinatorial entropy gain (`asymmetric' DNA), which results from the large number of possible binding partners for each DNA strand. We provide specific suggestions for the DNA sequences with which these effects can be achieved experimentally.

\end{abstract}

\maketitle

% PACS numbers: 05.70.Ce, 64.70.F-, 64.75.Cd, 02.70.Tt
(a) Electronic mail: bm411@cam.ac.uk, m.e.leunissen@amolf.nl, df246@cam.ac.uk

\section{Introduction}
The high selectivity of hybridisation of DNA makes it an interesting molecule to be used as  ``smart glue'' in the self-assembly of complex, nano-structured materials. Some of the advantages of DNA as a selective linker are that it can code for a large variety of specific linkages, it is readily available, it can be used under near ambient conditions and the binding process is reversible.  Consequently, applications of  DNA-mediated self-assembly range from computational biology \cite{DNAcomputingRev} and the assembly of scaffolded ``DNA-origami'' structures \cite{DNAorigami} to the development of targeting strategies (see  e.g.\ \cite{DH0DS0,sensor,MartinezPNAS}). Much experimental work has focused on the application of DNA as a selective linker that enables the self-assembly of  complex colloidal structures~\cite{DNA0a-1996,DNA0b-1996,GE-rev,GSE-2008,ExpRod2010,ExpRod2006,ExpNano2009,ExpNano2008,ExpNano2010,MLHSSM-2009,NMLG-2008,KSBSC-2008,Maye2010,Niemeyer2003,Biancaniello2005,Rogers2005,Valignat2005,Sun2005,Yao2007,Leunissen2009}. 
Several examples have been reported in the literature of (relatively simple) DNA-linked colloidal crystals, either  consisting of  nano colloids (see e.g.~\cite{ExpNano2009,ExpNano2008,ExpNano2010,MLHSSM-2009,NMLG-2008}) or  of micron--size colloids  with complementary DNA coating (see e.g.~\cite{ExpRod2006}). 
In parallel, theoretical investigations have provided insight into the factors that influence DNA-mediated self-assembly~\cite{Tkachenko,MartinezPRL,Starr2010,Scarlett2010,LF-2004}.

There are, however, some disadvantages associated with the use of DNA as a tool to link (nano)colloids. In particular, the very factors that lead to the specificity and reversibility of DNA linkages also cause the strength of DNA mediated interactions to depend strongly on the external conditions, such as the temperature or ionic strength \cite{MBF-2010,GE-rev,LF-2011,ExpRod2010,Jin}. This sensitivity results in an abrupt onset of aggregation as the temperature is lowered - a phenomenon that  can be problematic for self-assembly because it narrows the ``window'' of conditions within which reversible self-assembly of ordered structures is possible. If this window is missed,  self-assembly either does not take place at all or results  only in disordered aggregates. It is therefore important to explore possible approaches to combine the high selectivity of DNA-mediated interactions with a more gradual response to external conditions.

In the present paper, we use numerical simulations to demonstrate that `competing' DNA interactions can be used to create colloidal systems with a more gradual temperature response. Specifically, we consider a binary mixture of colloids ($X$ and $X'$) that, unlike most systems studied so far, are functionalised by a mixture of different DNA strands ($\al,\be$ and $\al',\be'$, respectively). These DNA sequences are chosen such that $\al$ can bind to both  $\al'$ and $\be'$ (and similarly $\al'$ to $\al$ and $\be$), but $\be$ cannot bind to $\be'$ (we will show that such DNA sequences can be readily designed). Importantly, the $\al$--$\be'$ and $\al'$--$\be$ linkages are weaker than those between $\al$ and $\al'$. As a result, $\al$--$\al'$ can form at a higher temperature than the linkages involving $\be$ or $\be'$. However, whereas a given $\al$, $\al'$ pair can form only one $\al$--$\al'$ linkage, it can form two weaker linkages ($\al$--$\be'$ and $\al'$--$\be$). This leads to a competition between the different types of linkages in which both their free energy of formation and combinatorial entropy effects,which depend on the number of different ways in which the linkages can form, play a role - as is characteristic for systems with multivalent binding \cite{KB-2003,MCW-1998,KGS-2000,Dreyfus2009}. Here, we investigate how the majority of the linkages can switch from one type to the other, not only as a function of the temperature and the difference in binding strength of the two linkage types, but also as a function of the surface coverage and length ratio of the DNA constructs.

The remainder of this paper is organised as follows: in Sec.\ \ref{SecModel} we introduce the model and the Monte Carlo algorithm. Sec.\ \ref{SecRes} reports our results. In Sec.\ \ref{SecCL} we consider the symmetric model, in which the length of the competing DNA constructs is the same ($L_\al=L_\be$). We show how the effective pair interaction depends on the hybridisation free energy of the DNA sticky ends and how the energetically driven switching from one linkage type to another broadens the association-dissociation transition of the particles; we rationalise these findings in the context of a Mean-Field model. In Sec.\ \ref{Sec2L} we consider the asymmetric model ($L_\al<L_\be$). We show how the length ratio of the different DNA constructs can be used to enhance the bond switching through combinatorial entropy effects. Finally, in Sec.\ \ref{SecSeq} we indicate how our approach could be implemented experimentally.

\begin{figure}
\vspace{1.cm}
\includegraphics[angle=0,scale=0.6]{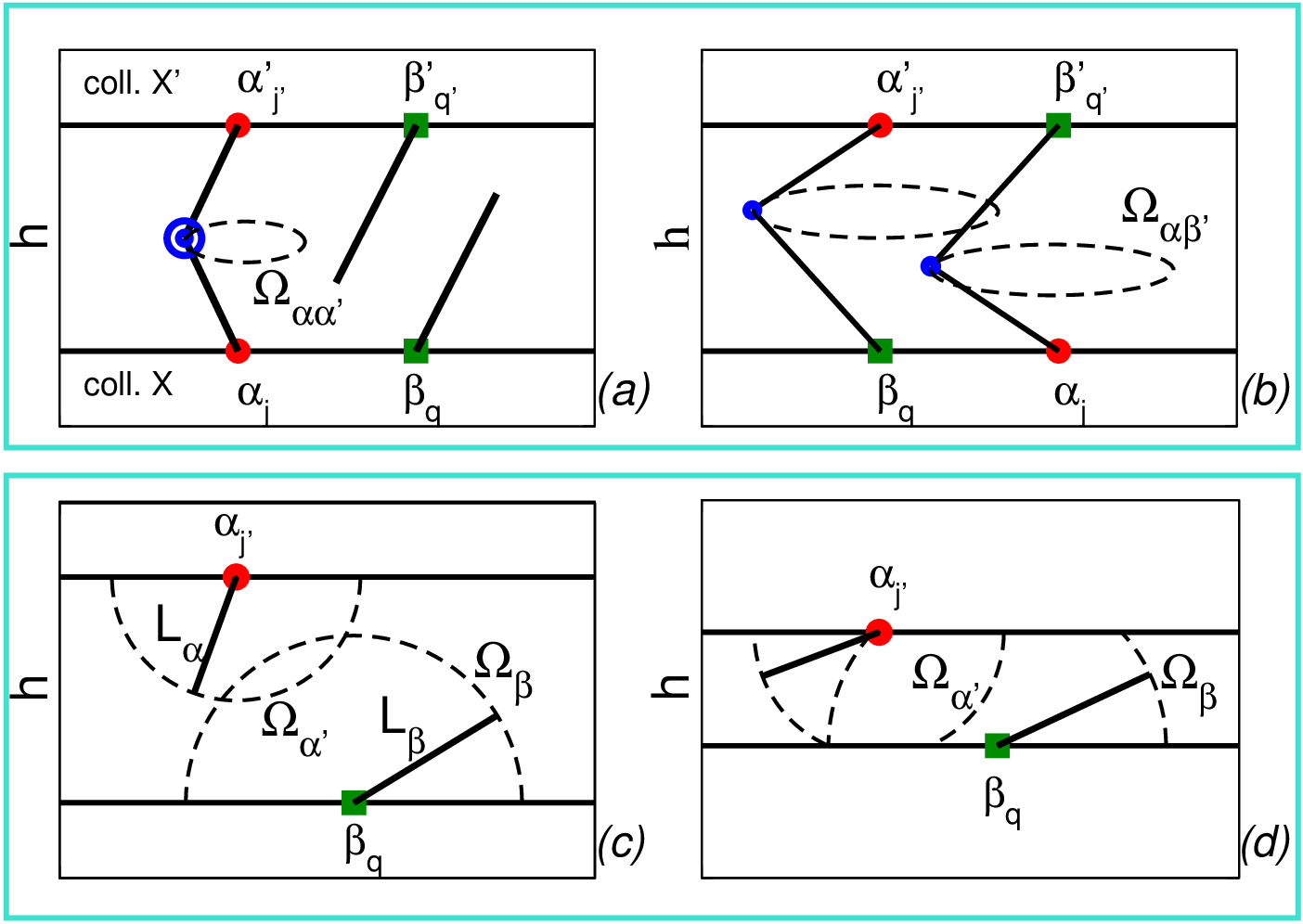} 
\vspace{1.1cm}
\caption{
Model system investigated in the present work. Pairs of colloids ($X$ and $X'$) are functionalised by two families of short DNA fragments ($\al$, $\be$ and $\al'$, $\be'$).  Colloids feature two kinds of linkages: $\al$--$\al'$ ({\em a}) and $\al$--$\be'$+$\al'$--$\be$ ({\em b}). When hybridised the conformational space of rods reduces to $\Omega_{\al\be'/\al\al'}$. Due to the excluded volume interaction between colloids and dsDNA, available configurations are also reduced when two colloids approach each other ({\em c} and {\em d}).
}\label{FigSys}
\end{figure}

\section{The Model}\label{SecModel}
 For the sake of simplicity, we use the relatively simple but well-tested  model of ref.~\cite{LF-2011} to describe the interaction between DNA-functionalised colloids.

Following ref.~\cite{LF-2011}, colloids are coated with double-stranded (ds) DNA that is shorter than its persistence length  of $\ell\approx 50\,\mathrm{nm}$ \cite{PersistenceLength} (corresponding to 150 nucleotides \cite{generalDNA}). The dsDNA is terminated by a short single-stranded (ss) DNA sequence ('sticky end') that can hybridise to complementary ssDNA. Binding is only allowed between colloids of type $X$ and $X'$. DNA  is modelled as thin rods (see Fig.\ \ref{FigSys}) randomly tethered to the colloidal surface. For typical surface coverages, the steric repulsion due to self-avoidance of the DNA strands ($2\,\mathrm{nm}$ diameter) is negligible compared to the entropic effects involved in the binding of tethered strands \cite{LF-2011} and we therefore treat the DNA as non-self-avoiding. Except for the DNA funtionalization, the colloids are assumed to be smooth and hard.  We consider two types of DNA ($\al$, $\be$ on $ X $ and $\al'$, $\be'$ on $ X '$, Fig.\ \ref{FigSys}) defined by different ssDNA end sequences, and (later on) different lengths $L_\al$, $L_\be$. The characteristic distance between DNA strands on a given colloid is $S=\sqrt{A/(N_\al+N_\be)}$, where $A$ is the total surface area of that colloid (say type $X$) and $N_\al$ and $N_\be$ are the total number of $\al$  and $\be$  strands on $ X $ (similar expressions apply to $ X '$).

Here we focus on the interaction between two parallel, planar surfaces. The pair potential between spherical colloids (of radius $R$) can be computed from these planar surface interactions via the Derjaguin approximation~\cite{Hunter}. In Ref.\ \cite{LF-2011} we have shown that this is a reliable approximation when $R/L\gtrsim 10$. From geometry, it further follows that the interactions are strictly pairwise additive if $R/L\gtrsim 6.5$, assuming that the hard cores of the particles can come into contact. For smaller particles, curvature and so-called three-body effects may cause deviations of the exact interactions from what is predicted here. Nevertheless, we expect that bond switching can also occur in these systems, albeit at different values for the relevant parameters.

In the simulations, we consider two square planes with side $L=\sqrt{A}$ and periodic boundary conditions in the directions parallel to the planes.  We fix $L_\al + L_\be=40\,\mathrm{nm}$ \cite{LF-2011}. In real units, the width of  the simulation box is $L\approx 0.5\, \mu\mathrm{m}$. Varying $N_\al$ and $N_\be$, we obtain characteristic inter-chain separations $S=0.53,0.75,1.06\cdot L_\al$. These values  span typical experimental values, e.g. \cite{ExpRod2010,ExpRod2006}. We verified that averaging  over different random realisations of tethering points does not alter our results within a 2\% of tolerance.

The  hybridisation of pairs of ssDNA depends on the specific nucleotide sequences of the individual strands and on the solvent properties (temperature and salt concentration). For the present paper it is important that ssDNA sequences can be designed in such a way that only the following linkages are possible: $\al$--$\al'$, but also $\al$--$\be'$ and $\al'$--$\be$ (see Fig.\ \ref{FigSys}{\em a} and {\em b}). In Sec.\ \ref{SecSeq} we show that it is possible to design nucleotide sequences that will yield this behaviour.

 We denote the hybridisation free energy of two free ssDNA by $\Delta G_{0x}$, where $x=\al$ for an $\al$--$\al'$ linkage and $x=\be$ or $x=\be'$ for an $\al$--$\be'$ or $\al'$--$\be$ linkage. We assume that $\Delta G_{0\al}<\Delta G_{0\be'}=\Delta G_{0\be}$. The binding free energy between two surface-bound DNA strands must also include a configurational entropy term \cite{LF-2011,Dreyfus2009}, which accounts for the reduced freedom of motion of the strands upon binding (Fig.\ \ref{FigSys}). If the dsDNA can swivel freely on the surface and the ssDNA is connected flexibly to the dsDNA \cite{ExpNano2008}, then the configurational space of the DNA rods  is a circle ($\Omega_{\al\al'}$ and $\Omega_{\al\be'}$ in Fig.\ \ref{FigSys}), that is bounded by the two surfaces. In App.\ \ref{AppTrunc} we give an explicit expression for $\Omega_{\al\be'}$ as a function of $L_\al$  and $L_\be$. As can be seen from Figs.\ \ref{FigSys}{\em c} and {\em d}, $\Omega_{\al}$ and $\Omega_{\be}$ are equal to $2\pi L_{\al/\be}^2$ if the distance between the planes $h$ is larger than $L_{\al/\be}$ and equal to $2\pi L_{\al/\be} h$ if $h<L_{\al/\be}$. As a result (see \cite{LF-2011} and App.\ \ref{AppDGcon}) the hybridisation free energy of two tethered strands $\Delta G_{\al x}$ ($x=\al'$ or $\be'$) is given by
\begin{eqnarray}
\DG_{\al x} = \Delta G_{0x} - k_B T \ln\Bigg({\tau\over \rho_0}{\Omega_{\al x} \over \Omega_\al \Omega_{x}}  \Bigg)
\quad ,
\label{eqDGcon}
\end{eqnarray}
where $\rho_0$ is the number density corresponding to one molar $\rho_0=6.022 \cdot 10^{23}/\mathrm{liter}$. $\tau$ is a non--universal factor that depends on the details of the linkage formed between the two dsDNA strands. As explained in App.\ \ref{AppDGcon}, slightly different expressions for $\tau$ result depending on how the sticky ends are coarse grained. However, the resulting differences in the predicted binding strengths are tiny and therefore  irrelevant for the present discussion.

In each MC move \cite{LF-2011}, we randomly select one of the $\al_j$ (or $\al'_{j'}$) functional arms and attempt to make, break or switch a linkage. First we consider the list of all the free $x'_{j'}$ ($x=\al,\be$) that can bind to $\al_j$, possibly including the partner to which $\al_j$ is already bound. A linkage $\al_j$--$x'_{q'}$ is created with probability $p_{q'}=\exp[-\beta \Delta G_{\al_j x'_{q'}}]/Q_j$. The probability that no linkage is formed is given by $p_0=1/Q_j$ , where  $Q_j=1+\sum_{x'_{q'}}\exp[-\beta \Delta G_{\al_j x'_{q'}}]$. $\Delta G_{\al x'}$ is computed using Eq.\ (\ref{eqDGcon}).  
In order to enhance the sampling  of the model we also implemented a `linkage-swapping' MC move that attempts to switch between a single $\al$--$\al'$ linkage and two weaker $\al$--$\be'$+$\al'$--$\be$ linkages. Details are given in App.\ \ref{AppMC}.

\begin{figure}
\vspace{0.5cm}
\includegraphics[angle=0,scale=0.5]{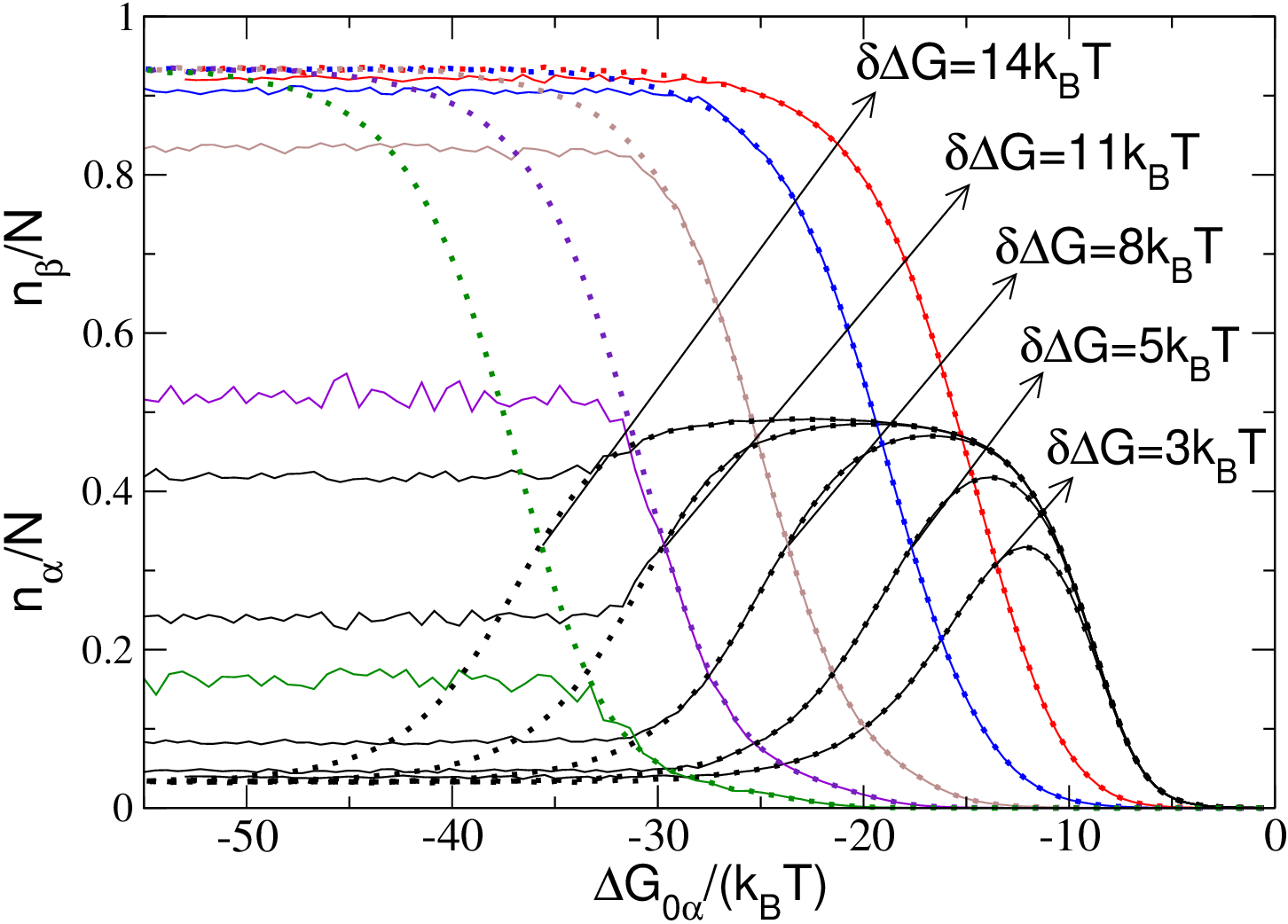} 
\vspace{0.5cm}
\caption{ 
Number of $\al$--$\al'$ linkages (black curves) and of $\al$--$\be'$+$\al'$--$\be$ linkages (coloured curves) versus $\Delta G_{0\al}$ (taken in unit of $k_B T$). In each simulation $\delta\Delta G= \Delta G_{0\be}-\Delta G_{0\al}$ is kept constant and equal to the value reported at the crossing point ($n_\al=n_\be$). Full lines refer to MC simulations that only use single bond rearrangements, while dotted lines refer to simulations that also employ `linkage--swapping' moves (App.\ \ref{AppMC}). We have used $N_\al = N_\be$ and $S=0.75L_\al$, while the distance between colloids is $h=L_\al=L_\be$.
}\label{FigInv}
\end{figure}

\section{Results}\label{SecRes}

\subsection{Symmetric DNA model ($L_\al = L_\be$)}\label{SecCL}

We first consider systems in which the competing DNA constructs have equal length $L_\al=L_\be$. Fig.\ \ref{FigInv} shows $n_\al$, the average number of $\al$--$\al'$ linkages, and $n_\be$, which denotes the number of $\al$--$\be'$+$\al'$--$\be$ linkages. $n_\al$ and $n_\be$ are plotted as a function of $\Delta G_{0\al}$. There is a linear relation between the hybridisation free energy and the temperature $\Delta G_{0\al}=\Delta H_{0\al}-T\Delta S_{0\al}$, where the hybridisation enthalpy/entropy ($\Delta H_{0\al}/\Delta S_{0\al}$) are negative constants. Hence decreasing $\Delta G_{0\al}$ corresponds to a decrease in $T$. For real DNA sequences, $\delta \Delta G\equiv \Delta G_{0\be}-\Delta G_{0\al}$ also depends on $T$ (see Sec.\ \ref{SecSeq}). However for the sake of simplicity in Fig.\ \ref{FigInv} we sketch our results at constant $\delta \Delta G$.

Upon decreasing $\Delta G_{0\al}$ from an initial situation where all DNAs are unbound, the stronger $\al$--$\al'$ linkages form first. For the specific choice of simulation parameters listed in Fig.\ \ref{FigInv}, this happens when $\Delta G_{0\al}\approx -5 k_B T$. As we lower $\Delta G_{0\al}$ even more,  the $n_\al$ linkages  disappear in favour of the weaker bonds between $\al$--$\be'$ and $\al'$--$\be$. This may seem counter-intuitive at first, but the reason is simple: replacing a single $\al$--$\al'$ linkage by two weaker ones ($\al$--$\be'$+$\al'$--$\be$) is favorable as long as the total gain in binding free energy upon forming these two linkages outweighs the loss in binding free energy when breaking an  $\al$--$\al'$ linkage. The switching between the different bond types in Fig.\ \ref{FigInv} is driven by the difference in the hybridization free energies of the individual linkages and the main effect of increasing $\delta \Delta G$ is to decrease the $\Delta G_{0\al}$ value of the crossing point where $n_\be = n_\al$. As will be shown below, the combinatorial entropy of the overall system \cite{Dreyfus2009,LF-2011} plays a less important role in this particular case.

\begin{figure}[h]
\vspace{1.5cm}
\includegraphics[angle=0,scale=0.3]{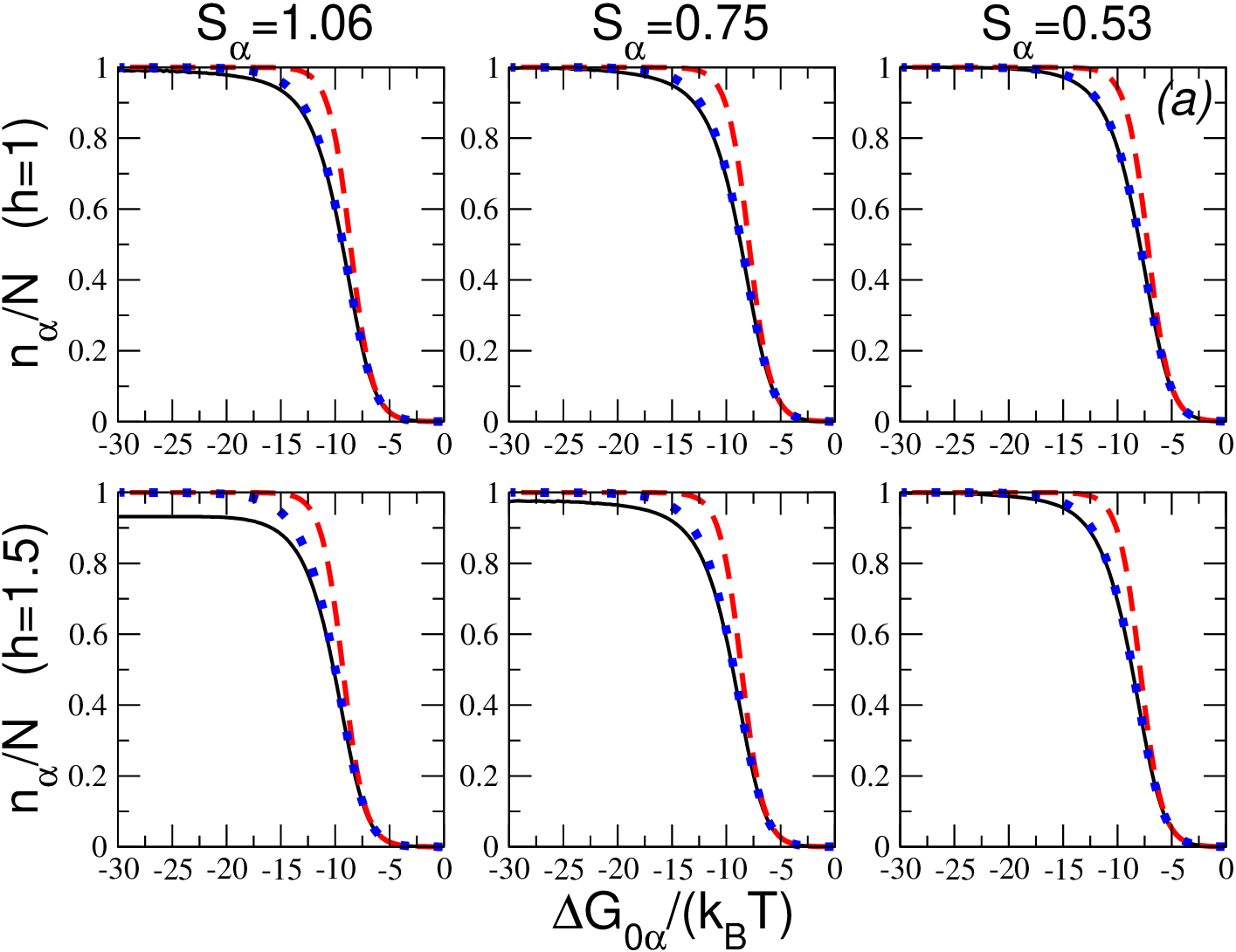}\quad
\includegraphics[angle=0,scale=0.3]{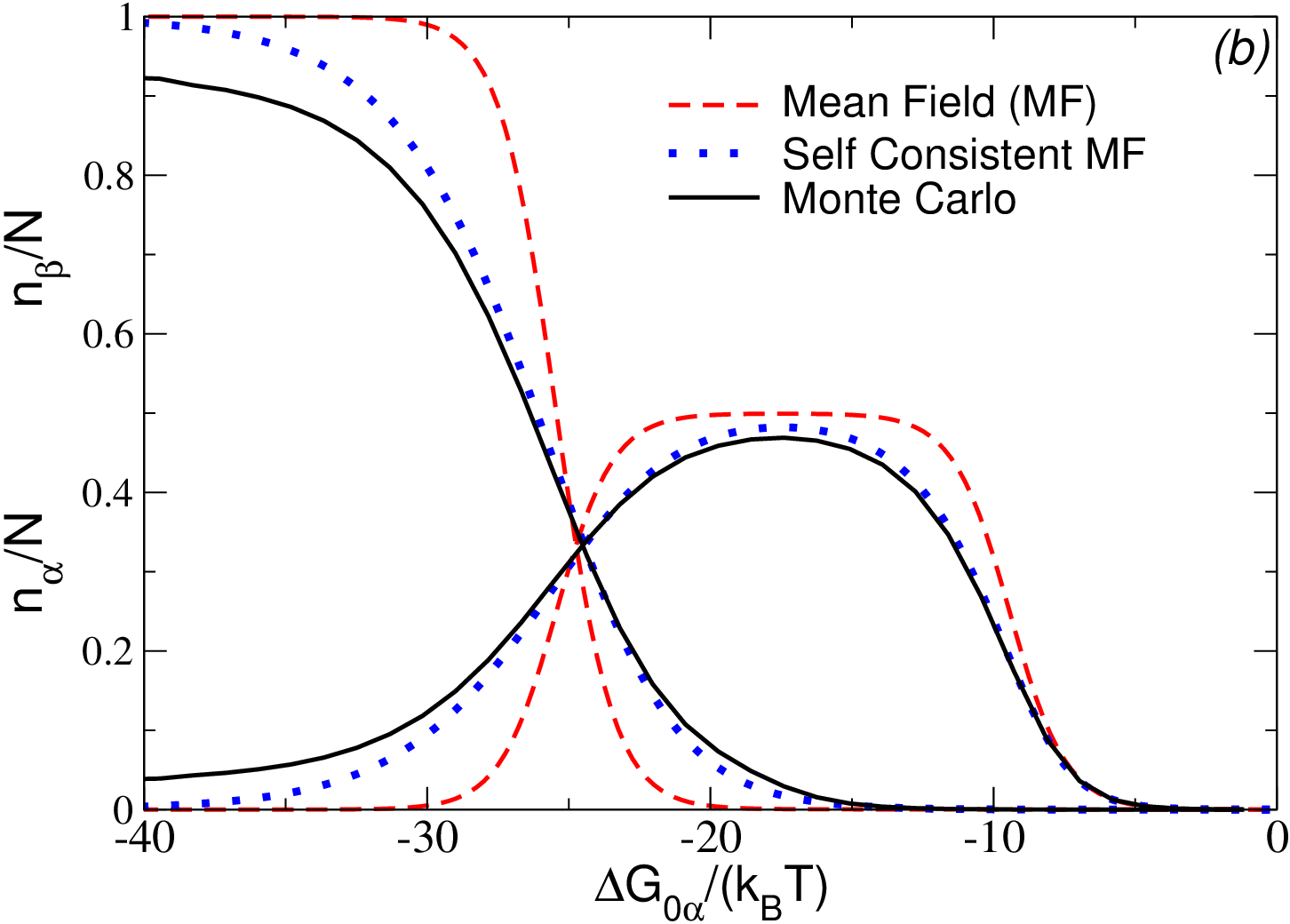}
\vspace{0.5cm}
\caption{ Monte Carlo simulations (full lines), Mean Field Eq.\ \ref{mean1} (broken lines), and self--consistent Mean Field Eq.\ \ref{mean2} results (dotted lines) for $n_\al$ and $n_\be$. Part ({\em a}) reports predictions for the single linkage model with $S_\al=1.06$ (first column), $S_\al=0.75$ (second column), $S_\al=0.53$ (third column), $h= L_\al$ (first row), and $h=1.5 L_\al$ (second row). Part ({\em b}) reports predictions for the competing linkages model with simulation parameters like in Fig.\ \ref{FigInv}.
}\label{figSC}
\end{figure}
The behaviour shown in Fig.\ \ref{FigInv} can be understood in a more quantitative way using a mean-field theory. We define $s_\al(h)$  and $s_\be(h)$ as the average configurational entropy cost of an $\al$--$\al'$ and $\al$--$\be'$ linkage
\begin{eqnarray}
s_\al(h) = {1\over A_{\al}}\int_{A_\al} \mathrm{d}{\bf x}' \, \exp\Big[ {\Delta S^{(\mathrm{cnf})}(L_\al,L_\al)\over k_B T} \Big] 
&\quad &
s_\be(h) = {1\over A_{\be}}\int_{A_\be} \mathrm{d}{\bf x}' \, \exp\Big[ {\Delta S^{(\mathrm{cnf})}(L_\al,L_\be)\over k_B T} \Big]
\quad ,
\nonumber\\
\label{mean0}
\end{eqnarray} 
where $A_\al$ and $A_\be$ are the circles on $ X '$ enclosing all the possible $\al'$ and $\be'$ (tethered at position ${\bf x}'$) which could be hybridised by $\al$ on $X$--DNA coated colloids represent a multivalent binding system in which each DNA strand can form several different linkages. In the low binding regime ($n_\al + n_\be/2 \ll N _\al$) this combinatorial entropy contribution is accounted for by noticing that the number of possible $\al$--$\al'$  and $\al$--$\be'$ linkages are $A_\al S_\al^{-2}$ and $A_\be S_\be^{-2}$, where $S_\al$ ($S_\be$) is the mean distance between $\al$ ($\be$) strands. It follows that the partition function $Z$ and the average fraction of the two linkages ($p_\al=n_\al/N$ $p_\be=n_\be/N$) is:
\begin{eqnarray}
 Z &=& \sum_{2n_\al+ n_\be \leq 2 N_\al} {N_\al\choose n_\al}{2 N_\al-2 n_\al\choose n_\be}  \Big({ s_{\al}(h) e^{-\be \Delta G_{0\al}} \over S_\al^2/A_\al }\Big)^{n_\al} \Big({s_{\be}(h)e^{-\beta \Delta G_{0\be}} \over S_\be^2/A_\be}\Big)^{n_\be} 
\nonumber\\
 &=& \Big[1+ { s_{\al}(h) e^{-\be \Delta G_{0\al}} \over S_\al^2/A_\al}  + 2 {s_{\be}(h)e^{-\be \Delta G_{0\be}} \over S_\be^2/A_\be} +\Big( {s_{\be}(h)e^{-\be \Delta G_{0\be}} \over S_\be^2/A_\be}\Big)^2\Big]^{N_\al} 
\label{this}
\\
 p_\al(S_\al,S_\be) &=& { s_{\al}(h) e^{-\be \Delta G_{0\al}} \over  Z^{1/N_\al} S_\al^2 /A_\al}
 \nonumber \\
p_\be(S_\al,S_\be) &=& { 2 s_{\be}(h)e^{-\be \Delta G_{0\be}} + 2 \Big( s_{\be}(h)e^{-\be \Delta G_{0\be}} \Big)^2/(S_\be^2/A_\be) \over Z^{1/N_\al} S_\be^2 /A_\be} \, .
\label{mean1}
\end{eqnarray}
Fig.\ \ref{figSC} compares the mean-field predictions based on  Eq.\ (\ref{mean1})  for the single linkage model \cite{LF-2011} (part {\em a}) and for the competing linkages model of Sec.\ \ref{SecModel} (part {\em b}) with the results of the MC simulations. There is qualitative, though not quantitative agreement between theory and simulations. 

 A better agreement is obtained when we improve our estimate of the combinatorial entropy terms. If $n_\al$ linkages are already present, the number of ways an extra linkage can be added is smaller than if $n_\al=0$. This is, because the mean distance between un--hybridised $\al'$ strands increases like $S_{\al}/\sqrt{1-n_{\al}/N}$ (and similarly for $S_{\be}$). Instead of exactly dealing with this correction we  used Eq.\ (\ref{mean1}), while correcting $S_\al$ and $S_\be$ by the average number of linkages ($n_\al$ and $n_\be$) which are then computed self--consistently in the following way 
\begin{eqnarray}
{n_\al\over N} =  p_\al \Big( {S_\al \over \sqrt{1-n_\al/N_\al}},
{S_\be \over \sqrt{1-n_\be/(2 N_\be)}} \Big)
&\quad &
{n_\be\over N} = p_\be \Big( {S_\al \over \sqrt{1-n_\al/N_\al}},
{S_\be \over \sqrt{1-n_\be/(2 N_\be)}} \Big) \, .
\nonumber\\
\label{mean2}
\end{eqnarray}
Dotted curves of Fig.\ \ref{figSC} show the predictions based on the solution of Eq.\ \ref{mean2}. As can be seen from this figure, the agreement between theory and simulation is now satisfactory. Although producing different profiles, Fig.\ \ref{figSC}b shows that the two different estimates of the combinatorial entropy (Eqs.\ \ref{mean1} and \ref{mean2}), place the bond switching transition (where $n_\al=n_\be$) at the same $\Delta G_{0\al}$. Indeed, using Eq.\ (\ref{mean1}) without any combinatorial prefactors (but allowing for each $\alpha$/$\alpha'$ strand only one $\alpha'/\alpha$ and one $\beta'/\beta$ partner) we find that $\Delta G_{0\al}$ at the bond switching transition decreases by $\lessapprox 3k_B T$. This indicates that the bond switching observed here is mainly an `energetic' effect, having to do with the different hybridization free energies of the individual bonds.

\begin{figure}[h]
\vspace{1.5cm}
\includegraphics[angle=0,scale=0.3]{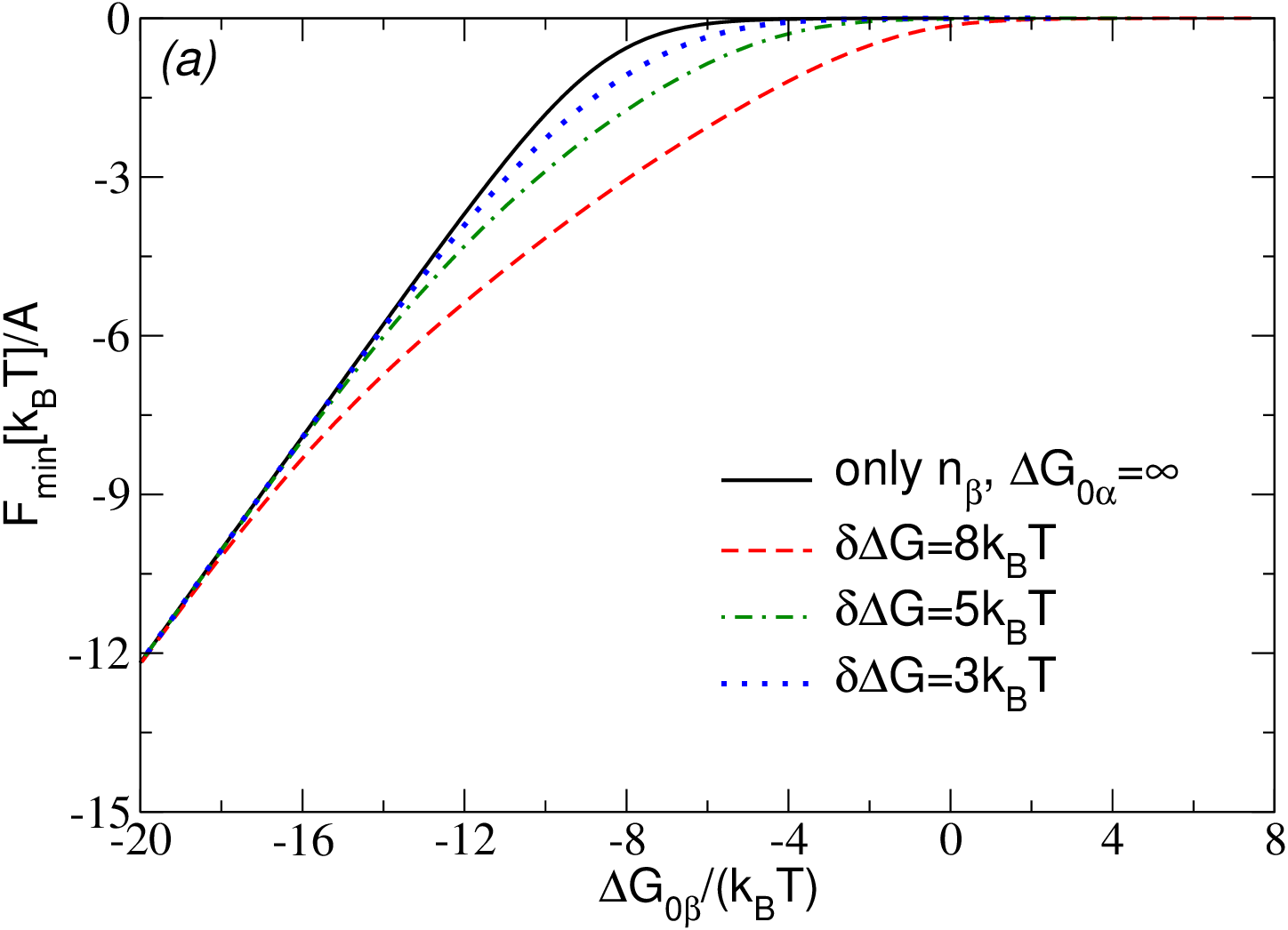}\quad
\includegraphics[angle=0,scale=0.3]{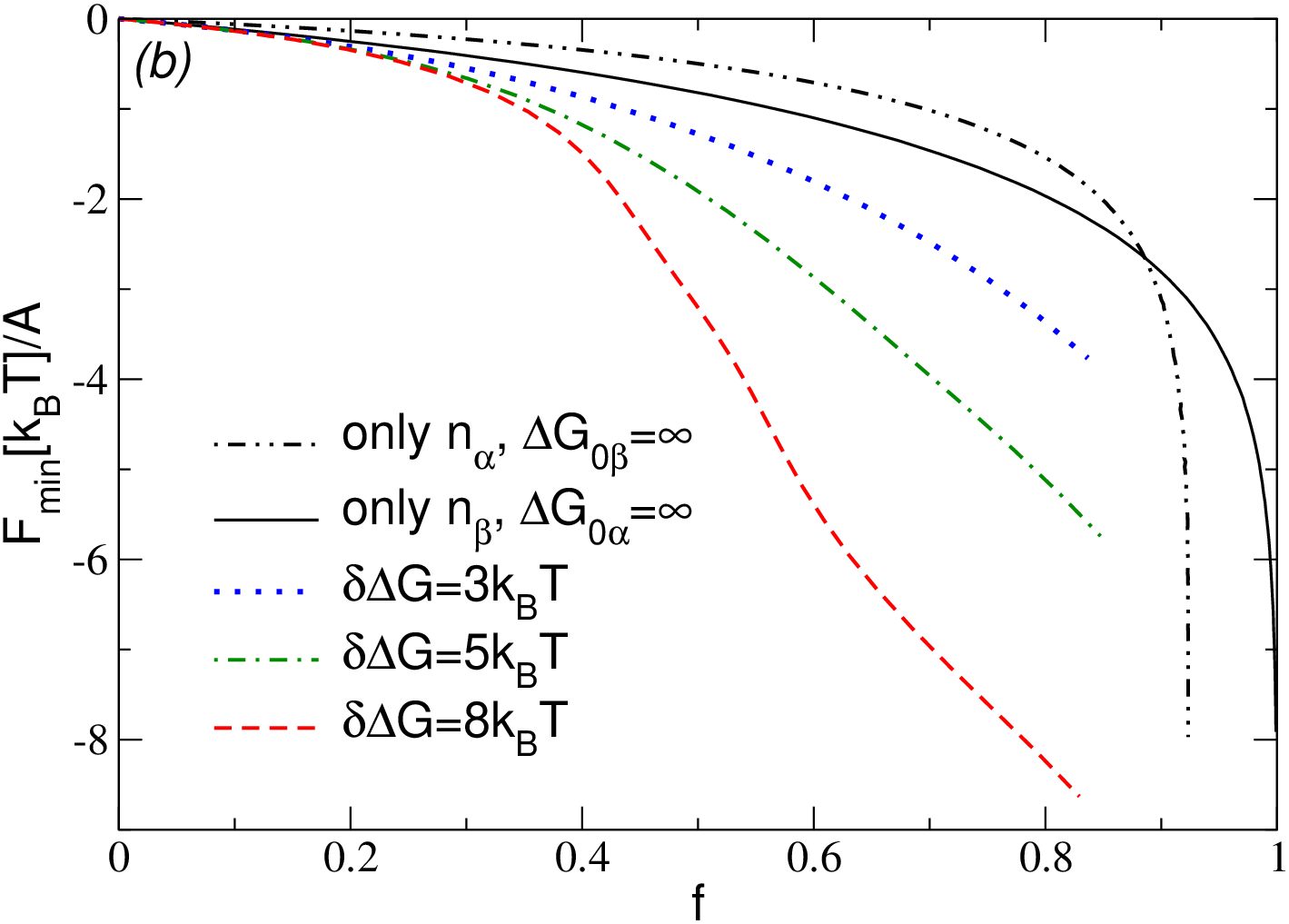}
\vspace{0.5cm}
\caption{ 
 Minima of the pair potential per unit area as a function of ({\em a}) $\Delta G_{0\be}$ and ({\em b}) the fraction of hybridised linkages. We report results for systems with only $n_\be$ or $n_\al$ linkages ($\Delta G_{0\al}=\infty$ or $\Delta G_{0\be}=\infty$), and for the competing linkages model with three values of $\delta \Delta G$. We have used $S=0.75\, L_\al$, $L_\al=L_\be=20\, \mathrm{nm}$, and  $N_\al/N_\be=3/7$.
}\label{MinimaPot}
\end{figure}

The fact that the dominant bond-type between colloids switches as the temperature is decreased, gives rise to a more gradual temperature dependence of the attractive interaction between the colloids. To see this, we consider the effective pair potential $F(h)$, where $h$ is the separation between the two surfaces. Following Ref.\ \cite{LF-2011}, $F(h)$ can be computed using thermodynamic integration starting from the case where the DNAs do not hybridise. In that case the pair potential is purely repulsive and is given per unit area by:
\begin{eqnarray} 
F_\mathrm{rep} (h) &=& {2 \over S_\al ^2}\mathrm{max}\Big[0,\ln{L_\al\over h}\Big] + {2\over S_\be^2} \mathrm{max}\Big[0,\ln{L_\be \over h}\Big] \, . 
\label{Vrep}
\end{eqnarray}
If we now consider colloids coated with two different types of linkers with constant $\delta \Delta G$, we can generalise the arguments of Ref.\ \cite{LF-2011} to show that the full potential is given by
\begin{eqnarray}
F (h) &=& F_\mathrm{rep}(h)+{1\over A}\int_{\infty}^{\Delta G_{0\al}} \mathrm{d} 
\Delta G_{0\al}' \,
< n_\al + n_\be>_{\Delta G_{0\al}',\delta  G}  \, .
\label{EqV}
\end{eqnarray}
$S_\al$ and $S_\be$ are the mean spacing of the $\al$ and $\be$ strands ($S_\al=\sqrt{A/N_\al}$, $S_\be=\sqrt{A/N_\be}$) and $A$ denotes the unit area inside which linkages are counted. In Eq.\ (\ref{EqV}) the average is taken with $\Delta G_{0\be} = \Delta G'_{0\al} +\delta \Delta G$. For the lower limit of the integration,  we choose a value for $\Delta G_{0\al}'$ such that $<n_\al+n_\be>\ll 1$. Fig.\ \ref{MinimaPot}{\em a} shows the dependence of the depth of the attractive well of $F(h)$ (usually located near $h=L$ due to the steric DNA--surface repulsion \cite{LF-2011}) as a function of $\Delta G_{0\be}$ for three different values of $\delta \Delta G$. For sake of comparison we also show the results for a system with the same distribution of $\al$ and $\be$ sticky ends, but in which the $\al$--$\al'$ linkages are forbidden (equivalent to taking $\Delta G_{0\al}=\infty$). This choice allows us to compare between a competing and a single linkage model with equal maximum number of possible bonds. At low $\Delta G_{0\be}$ only $n_\be$ linkages are present, and all the curves approach  the same binding strength. Importantly, the figure illustrates that the strength of the attraction changes more gradually with temperature when competing DNA linkers are present - this is especially evident at high $\delta \Delta G$. Experimentally, the more gradual response to temperature means that the DNA coated colloids display a broader association-dissociation transition and have a larger range of conditions under which they can form ordered assemblies (instead of kinetically disordered aggregates).

 Fig.\ \ref{MinimaPot}{\em b} shows $F_\mathrm{min}$ {\em versus} the fraction of hybridised bonds $f=(n_\al+n_\be)/N_\mathrm{tot}$, where $N_\mathrm{tot}$ is the maximum number of available linkages. Compared to single linkage models (black curves), the competing linkages model acquires a reasonable attractive well at lower $f$. It is worth remembering that the strong, high temperature $\al$--$\al'$ linkages are replaced with weaker $\al$--$\be'$ linkages as the temperature is lowered. The fact that the $\al$--$\al'$ linkages disappear before they become prohibitively strong means that kinetically trapped configurations should be automatically avoided, which represents a great experimental benefit. Indeed the formation of an 'irreversible' $\al$--$\al'$ linkage starting from two 'dynamic' $\al$--$\be'+\al'$--$\be$ ones needs the breaking of two independent linkages. Under bond--switching conditions, this requires an average time ($\Delta t\sim\exp[-2\beta\Delta G_{\al\be'}]$) even longer than the life--time of $\al$--$\al'$ ($\Delta t\sim\exp[-\beta \Delta G_{\al\al'}]$).
Nevertheless, Fig.\ \ref{FigInv} does show signs of equilibration problems. Using only sequential single linkage MC moves (full lines in Fig.\ \ref{FigInv}) we found that, at low $\Delta G_{0\al}$ and high $\delta \Delta G$, the system was not able  to completely remove the $\al$--$\al'$ linkages. The problem is due to the fact that although the formation of two weaker bonds is thermodynamically favourable the system needs to first break a strong linkage. To equilibrate the system (dotted lines in Fig.\ \ref{FigInv}) it was necessary to use the MC move described in App.\ \ref{AppMC}, which swaps between a strong linkage and two favourable weaker ones. This implies that, in order to engineer kinetically accessible experiments, it is important to design nucleotide sequences with a small $\delta \Delta G$, and to think about possible strategies to push the bond switching transition to higher $\Delta G_{0\al}$. This will be investigated in the next section.

\subsection{Asymmetric DNA model ($L_\al < L_\be$)}\label{Sec2L}

In the previous section we have discussed symmetric systems ($L_\al=L_\be$) for which the switching from one strong bond to two weak bonds was mainly driven by the gain in binding free energy, due to the difference in hybridization free energy of the different linkage types. However, the bond switching can be further enhanced and, more importantly, kinetically enabled, by modifying the number of possible binding partners that each type of linker `sees' on the opposing surface. In doing so, we take full advantage of the multivalent nature of the system, going from a mostly `energetically' driven bond switching mechanism to one that is driven by combinatorial entropy effects.

Eqs.\ \ref{mean1} show that multivalence controls the appearance of $\al$--$\al'$ or $\al$--$\be'$+$\al'$--$\be$ linkages by $A_\al/S_\al^2$ and $A_\be/S_\be^2$, where $A_\al$ and $A_\be$ is the area of the surface that enclose the tethering points of all the possible targets of $\al$ ($\al'$ and $\be'$), while $S_{\al/\be}$ is the mean distance between DNA strands. Using simple algebra we find
\begin{eqnarray}
{A_\al\over S_\al^2} ={4 \pi L_\al^2 N_\al \over A} \Big[1-
{h^2 \over (2 L_\al)^2 } 
\Big]
&\qquad &
{A_\be\over S_\be^2} = {\pi (L_\al +L_\be)^2 N_\be \over A} \Big[1-
{h^2 \over (L_\al+L_\be)^2 } 
\Big] \, ,
\label{CombEn}
\end{eqnarray}
where we have introduced the number of $\al/\be$ strands ($N_{\al/\be}$) for unit area $A$.
One obvious way to favour the weaker $\al$--$\be'$ over the stronger $\al$--$\al'$ bonds is to decrease the concentration of $\al$-terminated DNA ($N_\al$) relative to that of the more weakly binding $\be$-DNA ($N_\be$). However Eq.\ (\ref{CombEn}) suggests that  $\al$--$\al'$ linkages may be suppressed even more efficiently by decreasing the length $L_\al$. For instance, at $h=L_\al$, the same  effect of having a relative concentration of $\al$ strands equal to 1\%, can be obtained by an asymmetric system with $L_\al=1/3L_\be$.
\begin{figure}[h]
\vspace{1.5cm}
\includegraphics[angle=0,scale=0.3]{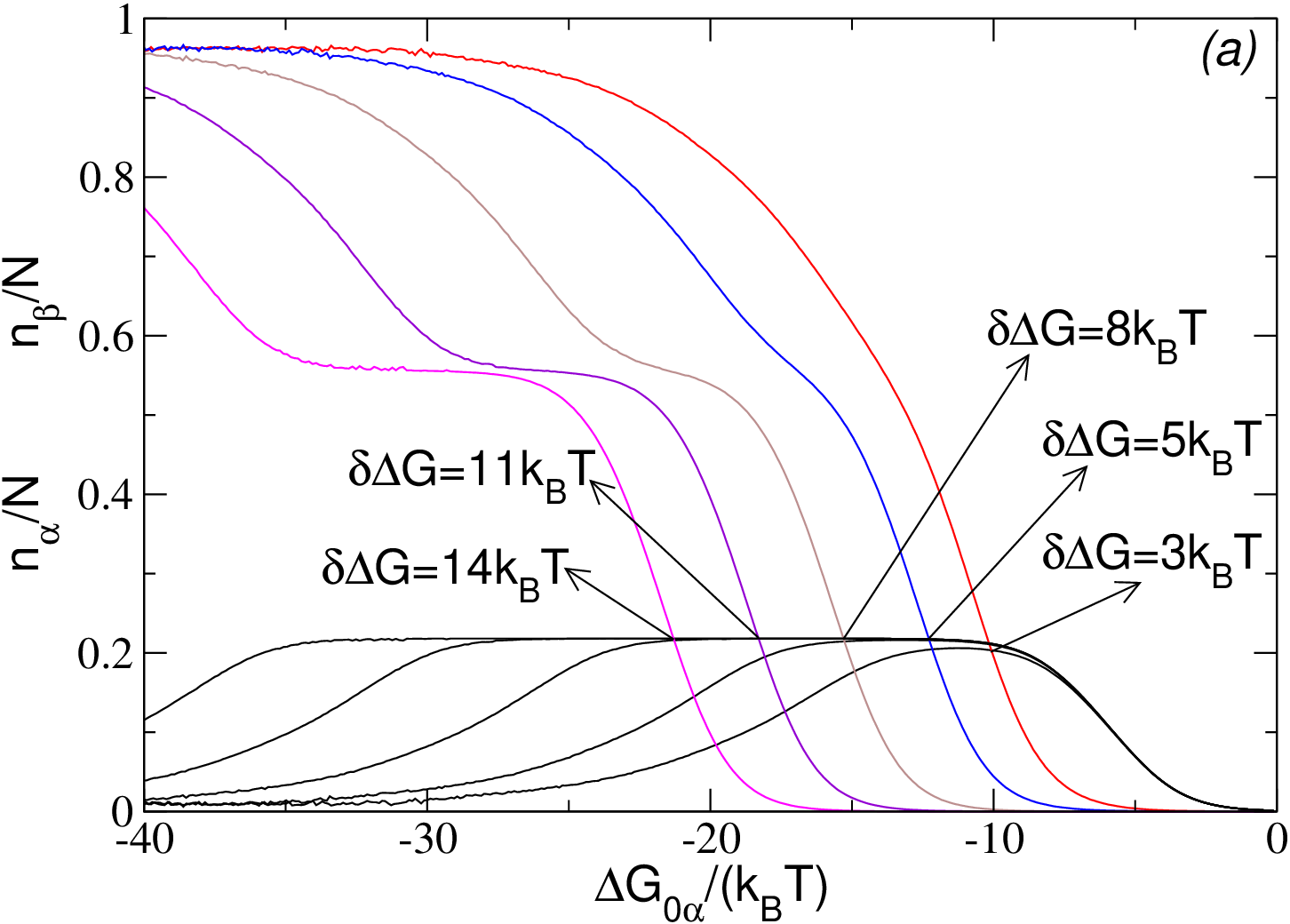}
\quad
\includegraphics[angle=0,scale=0.3]{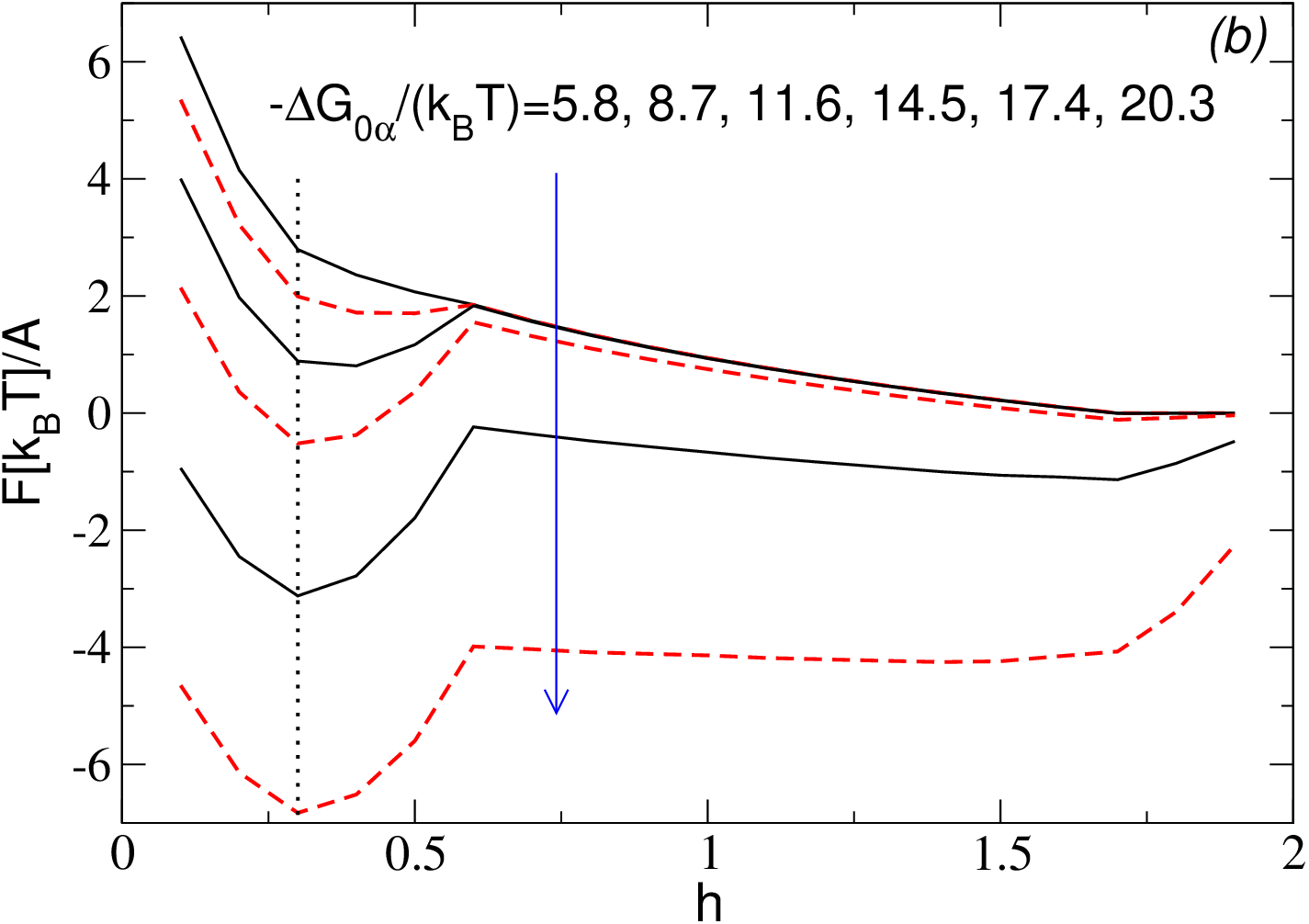}
\vspace{0.5cm}
\caption{({\em a}) Same as in Fig.\ \ref{FigInv}{\em a}  ($N_\al=N_\be$) except that $L_\al=0.3$, $L_\be=1.7$ and $h=0.3$. ({\em b}) Colloidal pair potentials $F(h)$ (Eq.\ \ref{EqV}) at six different $\Delta G_{0\al}$ and $\delta \Delta G=8 k_B T$. The vertical line is the distance at which plots of part ({\em a}) have been recorded.
}\label{fig2L}
\end{figure}
Fig.\ \ref{fig2L}{\em a} shows $n_\al$ and $n_\be$ under the same conditions as those considered in Fig.\ \ref{FigInv}, but using $L_\al/L_\be=3/17$ ($L_\al+L_\be=40\, \mathrm{nm}$). As compared to Fig.\ \ref{FigInv} the $\Delta G_{0\al}$ value for which $n_\al=n_\be$ is $\approx 3 k_B T$ higher if $\delta \Delta G=3 k_B T$, $\approx 8 k_B T$ higher if $\delta \Delta G=8 k_B T$, and $\approx 14 k_B T$ higher if $\delta \Delta G=14 k_B T$. 
We used both local and linkage swapping MC moves (App.\ \ref{AppMC}).
Interestingly, the $n_\al$ linkages first saturate at a fraction well below $0.5$ ($=N_\al / N$). This means that many $\al$ and $\al'$ sticky ends remain free and that the $n_\be$ linkages are formed much earlier than in the symmetric case (see Fig.\ \ref{FigInv}). The plateau value where $n_\al$ saturates depends on the DNA mean distance $S$ (Eq.\ \ref{CombEn}) and ranges from $0.14$ if $S=1.06$ to $0.31$ if $S=0.53$. The value of $\Delta G_{0\al}$ where $n_\al=n_\be$ also has a dependence on the coverage density, when using two lengths for the DNA constructs, but this dependence is found to be small ($\lessapprox 2 k_B T$ for the tested cases).

Fig.\ \ref{fig2L}{\em b} shows the effective pair potential per unit area at different values of $\Delta G_{0\al}$, keeping $\Delta G_{0\be} - \Delta G_{0 \al}$ fixed at $8 k_B T$. For large values of $\Delta G_{0\al}$, there is a local minimum in the potential at $h=L_\al$. This minimum becomes a global minimum as $\Delta G_{0\al}$ is lowered.  Important features of Fig.\ \ref{fig2L}{\em b} are the presence of repulsive tails that become shoulders at low $\Delta G_{0\al}$. By using  a Derjaguin approximation \cite{Hunter,LF-2011} it is possible to design sensible interactions that are useful to stabilise complex assemblies.

\begin{figure}[h]
\vspace{1.5cm}
\includegraphics[angle=0,scale=0.4]{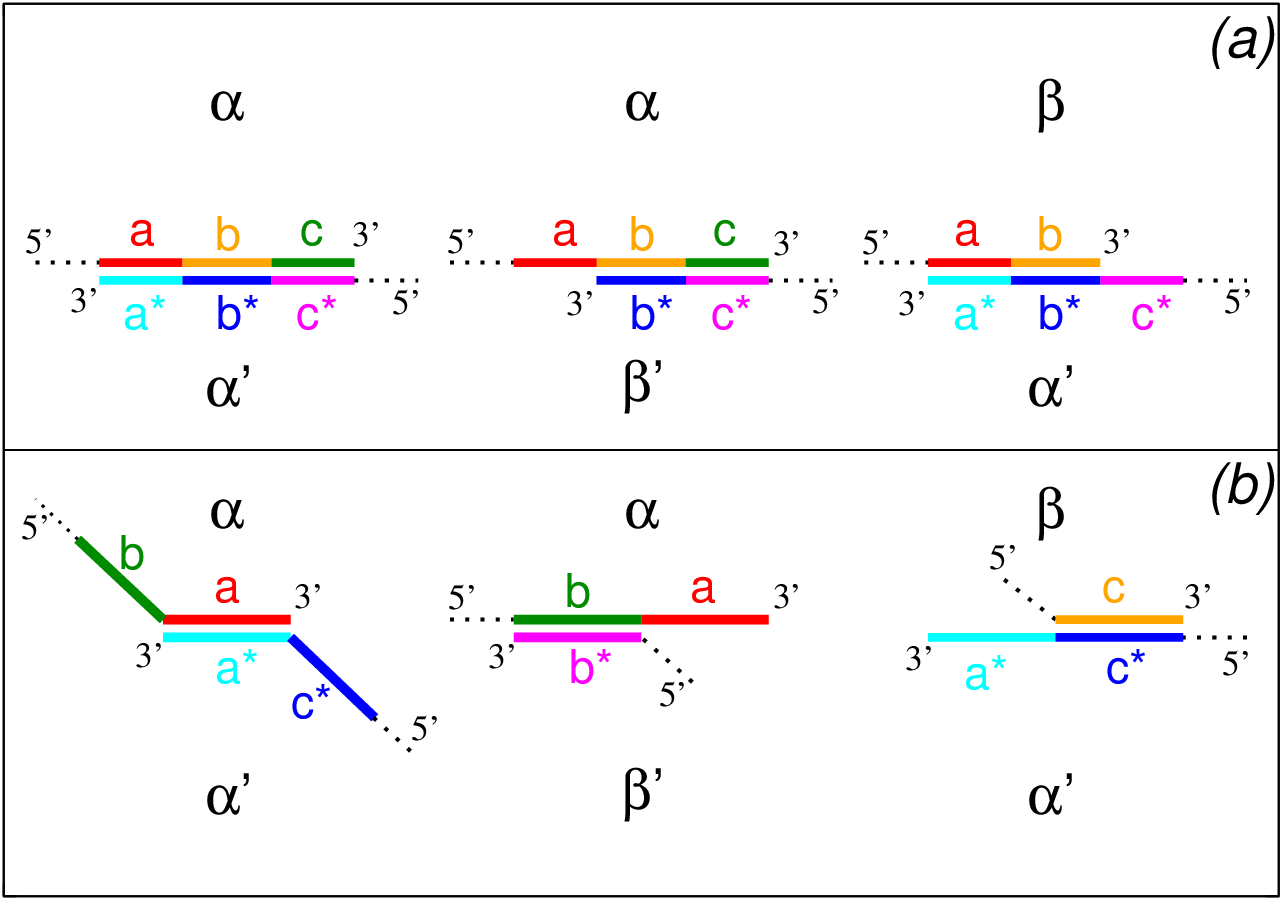}
\vspace{0.5cm}
\caption{ Two possible structures of the reactive end fragments (thick lines) with hybridised states that involve different subdomains (coloured segments labeled by $a$, $b$, and $c$). $a^*$, $b^*$ and $c^*$ are the complementary strands.
}\label{foldfig}
\end{figure}

\section{Designing the sticky--end sequences}\label{SecSeq}

The analysis described in the previous sections allows us to predict the relation between the formation of  $\al$--$\al'$,  $\al$--$\be'$ or $\al'$--$\be$ bonds and the hybridisation free energies of the individual  strands. In addition, we can now understand the effect of surface coverage and chain length. We find that  the dsDNA lengths ($L_\al$ and $L_\be$)  are the most relevant parameters that control the bond switching transition.

We now consider possible choices for the sticky end sequences that in experiments would result in a broadened dissociation transition of the colloids. We need to design our ssDNA sequences such that the $\al$ sequence binds more strongly to $\al'$ than to $\be'$ (and similarly for $\al'$). If we limit the discussion to  Watson-Crick base pairs, then different domains of $\al$ hybridise with $\al'$ and with $\be'$. In Fig.\ \ref{foldfig} we consider two architectures for the ssDNA in which the $\al$--$\al'$  linkage involves ({\em a}) the full length of the reactive fragments or ({\em b}) pieces of ssDNA which are not used in the weak bonds. As a proof of concept in Tab.\ \ref{TableSeq1}, we report possible sequences for these two families (A for Fig.\ \ref{foldfig}{\em a}  and B for Fig.\ \ref{foldfig}{\em b}). 

\begin{table}[h]
\begin{tabular}{|l|rcl|rcl|}
\hline
seq.\ $A_1(n)$ & $\al$ &=& $5'-[C_n][T_n][C_n]-3'$ & $\al'$ &=& $5'-[G_n][A_n][G_n]-3'$ \\
 & $\be$ &=& $5'-[C_n][T_n]-3'$ & $\be'$ &=& $5'-[G_n][A_n]-3'$ \\
\hline
seq.\ $A_2(n)$ & $\al$ &=& $5'-[T_n][C_n][T_n]-3'$ & $\al'$ &=& $5'-[A_n][G_n][A_n]-3'$ \\
 & $\be$ &=& $5'-[T_n][C_n]-3'$ & $\be'$ &=& $5'-[A_n][G_n]-3'$ \\
\hline
\hline
seq.\ $B_1(n)$  & $\al$ &=& $5'-[(TCT)_n][(GTG)_n]-3'$ & $\al'$ &=& $5'-[(ACA)_n][(CAC)_n]-3'$ \\
 & $\be$ &=& $5'-[(TGT)_n]-3'$ & $\be'$ &=& $5'-[(AGA)_n]-3'$ \\
\hline
seq.\ $B_2(n)$ & $\al$ &=& $5'-[TTGAGAAATCC][C_n]-3'$ & $\al'$ &=& $5'-[G_n][GGATCAATCTT]-3'$ \\
 & $\be$ &=& $5'-[AAGATTGATCC]-3'$ & $\be'$ &=& $5'-[GGATTTCTCAA]-3'$ \\
\hline 
\end{tabular}
\caption{
 Nucleotide sequences giving rise to the hybridised states reported in Figs.\ \ref{foldfig}{\em a} and \ref{foldfig}{\em b} ($B_2$ partially taken from \cite{BibSequences}). Brackets group different subdomains ($a$, $b$ and $c$) as defined in Fig.\ \ref{foldfig}, while subscripts stand for repeated nucleotides. 
}
\label{TableSeq1}
\end{table}

Sequences of type $A$ have been chosen with $a=c$ (Fig.\ \ref{foldfig}{\em a}). This straightforwardly balances the hybridisation free energy of the $\al$--$\be'$ and the $\al'$--$\be$ linkages \cite{NearestNeighbour}. The bottleneck is the possibility of $\be$--$\be'$ linkages, weaker than $\al$--$\be'$. This problem is avoided by using the scheme of Fig.\ \ref{foldfig}{\em b} and less degenerate sequences ($B_1$ and $B_2$) to further enforce the selectivity of each subdomain. In particular $B_2$ has been assembled using for the weak linkages two couples of sequences with nearly equal hybridisation free energy \cite{BibSequences}.

\begin{figure}[h]
\vspace{1.5cm}
\includegraphics[angle=0,scale=0.5]{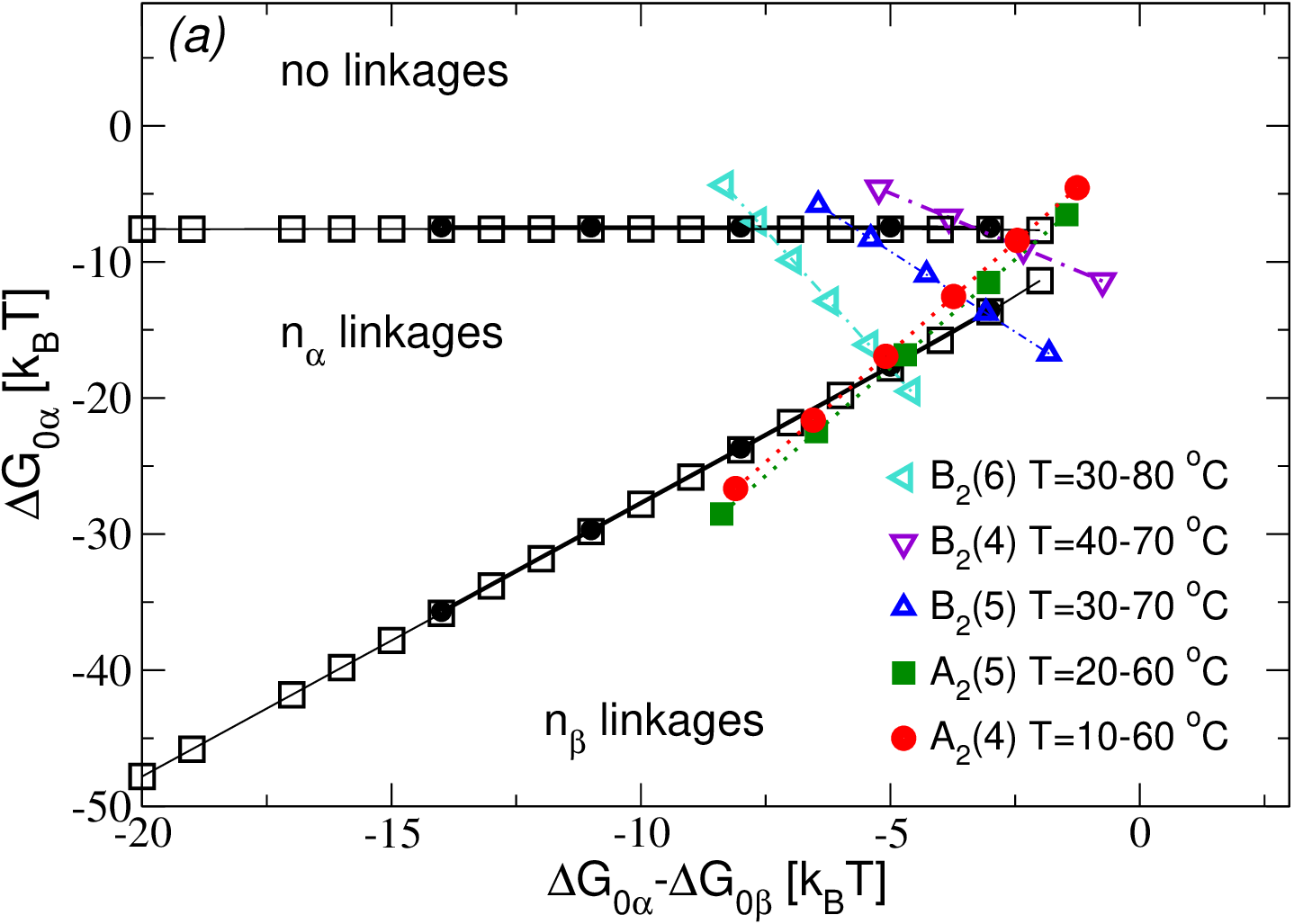} \\
\vspace{1.5cm}
\includegraphics[angle=0,scale=0.5]{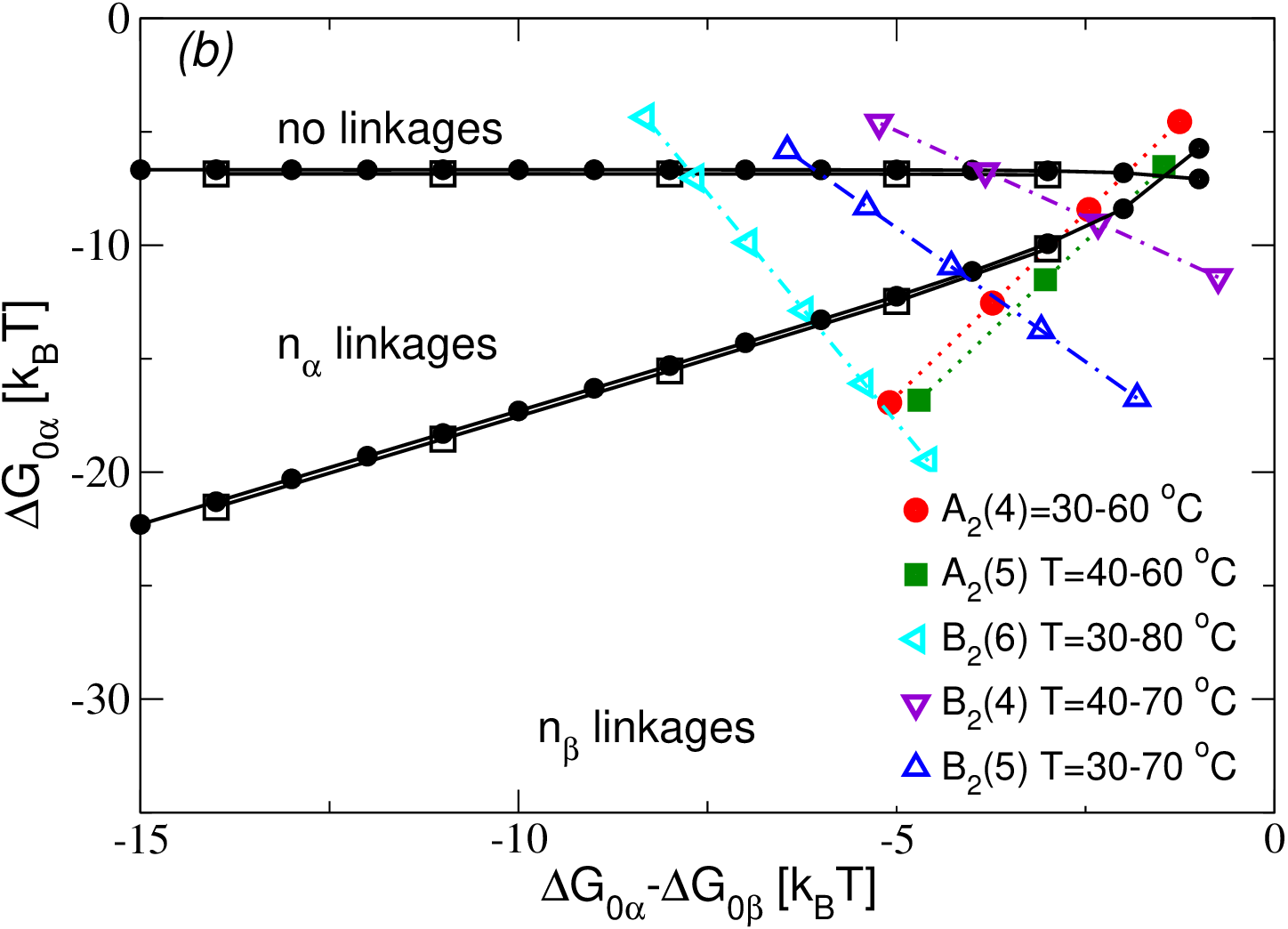} 
\vspace{0.5cm}
\caption{ 
Full black lines are MC results which outline the kind of linkages (none, $\al$--$\al'$, or $\al$--$\be'$) in the $\{\Delta G_{0\al},\Delta G_{0\al}-\Delta G_{0\be}\}$ plane. Full circles and open squares differ in the way the reactive sticky ends are coarse grained (see App.\ \ref{AppDGcon}). Part ({\em a}) refers to the symmetric system ($L_\al=L_\be$, $h=L_\al$), while part ({\em b}) to $L_\al=3/17 L_\be$, $h=L_\al$. Broken lines list some sequences of Tab.\ \ref{TableSeq1} within a temperature range (recorded every $10\, ^{\circ}C$) where the crossover between different linkages is expected to happen.
}\label{FigSequences}
\end{figure}
 As input for our model we use the hybridisation free energies of ssDNA fragments in solution. We used the ``DINAMelt'  estimates of  the ssDNA binding free energies~\cite{DINAMeltRef}.  For more details on the procedure, see Appendix~\ref{AppDGcon}. Fig.\ \ref{FigSequences} shows the relation between  $\Delta G_{0\al}$ and $\Delta G_{0\al}-\Delta G_{0\be}$ that  is predicted for $A_2(n)$ and $B_2(n)$ listed in Tab.\ \ref{TableSeq1} and different values of $n$. Based on the MC results of the previous section, we can predict the different bonding regimes both for DNA strands of equal length (Fig.\ \ref{FigSequences}{\em a}) and for the asymmetric case $L_\al/L_\be=3/17$ (Fig.\ \ref{FigSequences}{\em b}).

 Fig.\ \ref{FigSequences} can be used as the starting point to design possible experiments. First, given a certain colloidal architecture, MC results allow to draw the region where none, $n_\al$ or $n_\be$ linkages are expected. For given ssDNA sequences, it is then possible to predict the transition temperatures computing $\Delta G_{0\al}$ and $\delta\Delta G$ and overlapping them with MC results like those in Fig.\ \ref{FigSequences}. Sequences can be optimised (e.g.\ changing $n$ in Tab.\ \ref{TableSeq1}) to avoid kinetically trapped configurations.

\section{Conclusions}\label{SecConclusions}
In this paper we have proposed and tested a strategy that could make the crystallization transition of DNA-functionalised colloids less sensitive to external conditions. 

Specifically, we have considered a binary system of colloids ($X$ and $X'$) covered by two families of reactive sticky ends ($\al$, $\be$ on $X$ and $\al'$, $\be'$ on $X'$). Exploiting the selectivity of DNA, we have shown that it is possible to design sequences that only allow for binding between $\al$--$\al'$, $\al$--$\be'$ and $\al'$--$\be$. We choose the hybridisation free energy of $\al$ and $\al'$ in solutions stronger than the other two possible pairings with equal hybridisation free energy. Because $\al$ or $\al'$ participate in all the possible linkages, the strong $\al$--$\al'$ linkages compete with the weaker ones ($\al$--$\be'$+$\al'$--$\be$). However, the number of possible weaker linkages is twice the possible number of the strong ones.

We have demonstrated that a bond switching transition is possible: upon decreasing the temperature the first linkages to appear are the strongest $\al$--$\al'$ ones while at lower temperatures the system can gain free energy by replacing $\al$--$\al'$  with $\al$--$\be'$+$\al'$--$\be$. For symmetric systems (in which the length of the DNA constructs is equal), the bond switching is energetically driven. This means that the hybridisation free energy of having two weaker linkages is lower than the hybridisation free energy of a single strong one. When choosing the length of the $\al$ and $\al'$ linkers shorter than $\be$ and $\be'$ (asymmetric model) the bond switching transition is enhanced and occurs at higher temperatures than in the symmetric case. Here the transition is driven by the combinatorial entropy gain related to the fact that an $\al$ strand, for instance, can bind more $\be'$ than $\al'$.

The main effect of the competing DNA linkages is that the resulting effective inter-colloid pair potential is less strongly temperature dependent than is observed in the conventional case where only a single type of linkage is possible. This enhances the experimental control over the self--assembly of DNA functionalized colloids. A further advantage of the proposed strategy is that the strong linkages are replaced as temperature is lowered, which could be used in in step--wise assembly schemes.

Our procedure predicts the temperature range where transitions between different kinds of linkages are to be expected in experiments. Moreover, we have shown how one can optimise the DNA sequences and colloid architectures such that the linkage switching transition remains kinetically accessible.

{\bf ACKNOWLEDGMENTS:} This work was supported by the ERC (Advanced Grant agreement 227758). DF acknowledges support from a grant of the Royal Society of London (Wolfson Merit Award). The work of ML is part of the research programme of the Foundation for Fundamental Research on Matter (FOM), which is part of the Netherlands Organisation for Scientific Research (NWO). We acknowledge S.\ Angioletti-Uberti for a critical reading of the manuscript; M.\ S.\ Llewellyn-Jones, B.\ M.\ Mladek, and F.\ J.\ Martinez-Verachocea for useful discussions.

\appendix

\begin{figure}
\vspace{0.5cm}
\includegraphics[angle=0,scale=0.4]{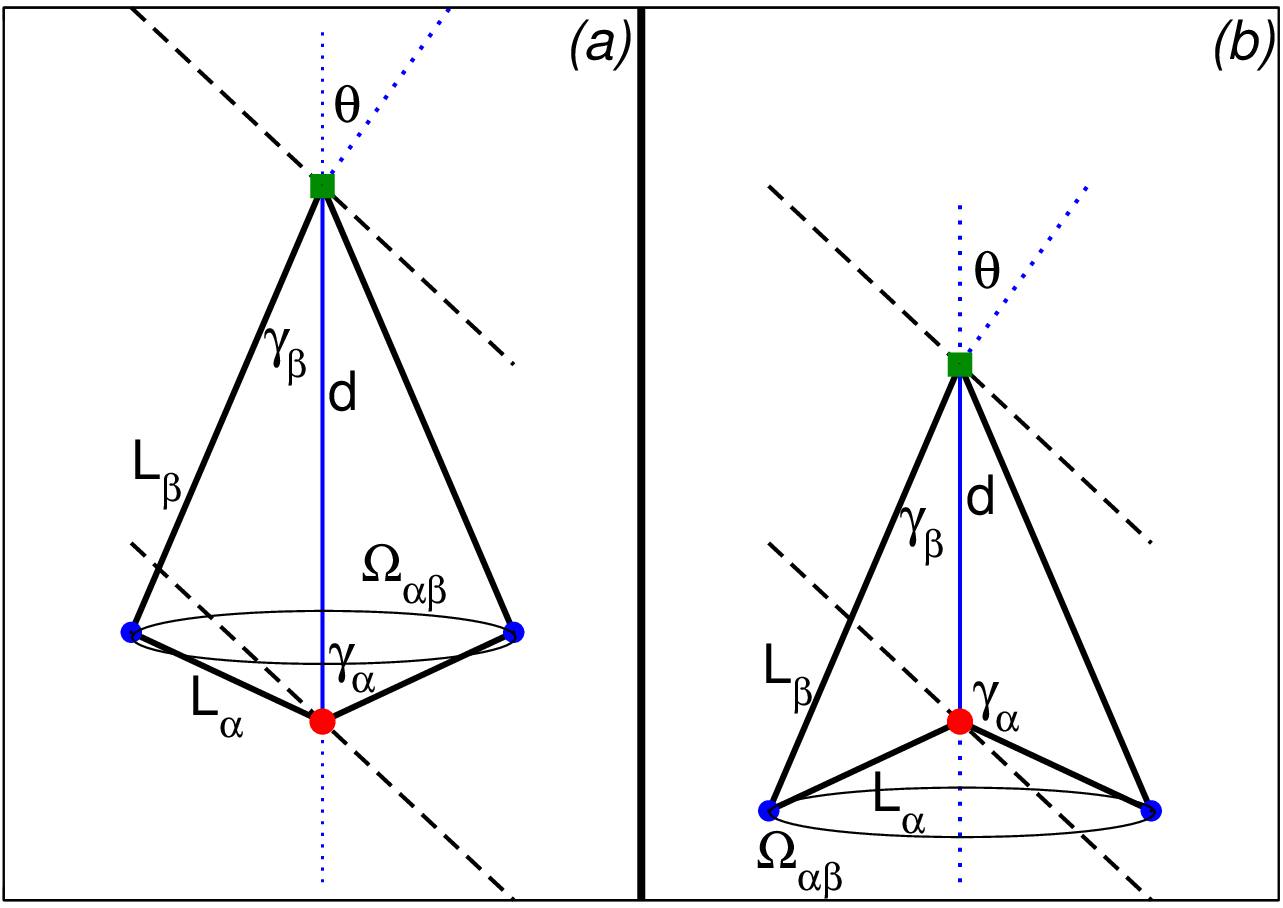} 
\vspace{0.5cm}
\caption{ Two possible ways by which the colloids (broken lines) can cut the configurational space of the hybridised rods $\Omega_{\al\be}$. Note that we choose a reference frame with the distance between the surface attachment points of the sticky ends $d$ taken along the vertical direction. 
}\label{figAppA}
\end{figure}

\section{Computation of $\Omega_{\al\be}$}\label{AppTrunc}

Here, we derive expressions for the configurational space of the hybridised strands, $\Omega_{\al\be}$, (Eq.\ \ref{eqDGcon}) as a function of the distance between the tethering points of the hybridised sticky ends $d$, the rods lengths $L_\al$, $L_\be$ ($L_\al<L_\be$), and the angle $\theta$ between ${\bf d}$ and the vector normal to the colloid surface (see Fig.\ \ref{figAppA}). If $\gamma_\al$ is the angle\footnote{$\gamma_\al$ is given by $\cos \gamma_\al=(d^2+L_\al^2 - 2 d L_\be)/2$. A similar relation holds for $\gamma_\be$.} between $L_\al$ and ${\bf d}$, depending on the colloidal distance and the rod lengths, it may be $\gamma_\al<\pi/2$ or $\gamma_\al>\pi/2$. In the first case (Fig.\ \ref{figAppA}{\em a}) $\Omega_{\al\be}$ may have none (Eq.\ \ref{Anone}), one (Eq.\ \ref{Aone}) or two cuts (Eq.\ \ref{Atwo})
\begin{eqnarray}
&& \Omega_{\al\be}= 2 \pi L_\al \sin\gamma_\al \qquad 
 \mathrm{if} \quad \theta < {\pi\over 2} - \gamma_\al 
\label{Anone}
\\
&& \Omega_{\al\be}=[2 \pi -\Delta\varphi(\gamma_\al,\theta) ] L_\al  \sin 
\gamma_\al 
\qquad 
 \mathrm{if} \quad {\pi\over 2} -\gamma_\al <\theta < {\pi\over 2} - \gamma_\be 
\label{Aone}
\\
&& \Omega_{\al\be}=[2 \pi -\Delta\varphi(\gamma_\al,\theta)-\Delta\varphi(\gamma_\be,\theta)] L_\al  \sin \gamma_\al \qquad 
 \mathrm{if} \quad {\pi\over 2} -\gamma_\be <\theta 
\label{Atwo}
\end{eqnarray}
where we have defined $\Delta\varphi (\gamma,\theta)$ as the planar angle of a cone (of amplitude $\gamma$) which is cut by a plane tilted by an angle $\theta$
\begin{eqnarray}
\Delta\varphi(\gamma,\theta) &=& 2\, 
 \mathrm{ArcCos} \left({1 \over \tan \gamma \tan\theta}\right) \quad .
\nonumber
\end{eqnarray}
If $\gamma_\al>\pi/2$ (Fig.\ \ref{figAppA}{\em b}) there might be a single cut by the plane where $L_\al$ is tethered (Eq.\ \ref{A1one}), or the second plane could also cut $\Omega_{\al\be}$ (Eq.\ \ref{A1two})
\begin{eqnarray}
&& \Omega_{\al\be} =  \Delta\varphi(\pi-\gamma_\al,\theta) L_\al \sin \gamma_\al
\qquad \mathrm{if} \quad \theta > \gamma_\al-{\pi\over 2}\quad \mathrm{and}\quad
\theta < {\pi\over 2} -\gamma_\be
\label{A1one} \\
&& \Omega_{\al\be} =  [\Delta\varphi(\pi-\gamma_\al,\theta)-\Delta\varphi (\gamma_\be,\theta)] L_\al \sin \gamma_\al
\qquad \mathrm{if} \quad \theta > \gamma_\al-{\pi\over 2}\quad \mathrm{and}\quad
\theta > {\pi \over 2} -\gamma_\be \, . 
\nonumber \\
\label{A1two}
\end{eqnarray}
Notice that $\Omega_{\al\be}=0$ when $\theta<\gamma_\al-\pi/2$.  This happens only if $L_\al<L_\be$, implying that for the hybridization of asymmetric linkages $\theta$ needs to stay inside a narrower region than when $L_\al=L_\be$.

\section{Estimate of the configurational entropy}\label{AppDGcon}

 Following \cite{LF-2011}, we first consider ssDNA fragments in solution ($Y$, $Z$ and $YZ$) in thermodynamic equilibrium $Y+Z\rightleftharpoons YZ$. In the ideal limit, the partition functions of the three species are
\begin{eqnarray}
Q^f_Y = {1\over N_Y!}\Big( {z_Y V  \over  \Lambda_Y^3 } \Big)^{N_Y} 
& \qquad  Q^f_Z= {1\over N_Z!}\Big( {z_Z V  \over  \Lambda_Z^3 } \Big)^{N_Z} 
\qquad &
Q^f_{YZ}={1\over N_{YZ}!} \Big({z_{YZ} V\over  \Lambda_{Y}^3  \Lambda_Z^3} v_0\Big)^{N_{YZ}}  \, .
\label{conf1}
\end{eqnarray}
$z_Y$, $z_Z$ and $z_{YZ}$ are the contributions to the partition function due to the internal degrees of freedom \cite{LF-2011}, $ \Lambda_{Y/Z}$ are the de-Broglie wave lengths, $V$ and $N_{Y/Z/YZ}$ the volume and the number of molecules, while $v_0$ is related to the specific potential ($V_\mathrm{bond}$) by which the two fragments are hybridised. Using classic pair potentials would give, for instance, $v_0=(2\pi k_B T/k_H)^{3/2}$ in the case of a harmonic potential with spring constant $k_H$, and $v_0=4/3 \pi w^3$ for a square-like potential with amplitude $w$. At equilibrium the relation $\mu_{YZ}=\mu_Y+\mu_Z$ links the ratio of the internal partition functions ($z_{Y/Z/YZ}$) with the number densities ($\rho_{Y/Z/YZ}$) which can be expressed in terms of the equilibrium constant $K=\exp(-\beta \Delta G^0_{YZ})$ 
\begin{eqnarray}
{z_{YZ}v_0\over z_Y z_Z}={\rho_{YZ}\over \rho_Y \rho_Z} = {1\over \rho_0} e^{-\beta \Delta G^0_{YZ}} \quad .
\label{DG0}
\end{eqnarray}

We now consider the case of tethered rods. When un--hybridised the partition functions of the free fragments are $Q_Y= \Omega_Y z_Y /  \Lambda_Y^2$ and $Q_Z= \Omega_Z z_Z /  \Lambda_Z^2$, where $\Omega_{Y/Z}$ is the configurational space available to the sticky ends (Fig.\ \ref{FigSys}). When hybridised  
\begin{eqnarray}
Q_{YZ} &=& {z_{YZ}\over  \Lambda_Y^2  \Lambda_Z^2} \int_{\Omega_1} \di {\bf x}_1 \int_{\Omega_2} \di {\bf x}_2 \exp[-\beta V_\mathrm{bond}({\bf x}_1 - {\bf x}_2)] \quad ,
\label{intsph}
\end{eqnarray}
where ${\bf x}_1$ and ${\bf x}_2$ are the coordinates of the two sticky ends. Taking the large $k_H$ limit (for spring potentials) or the small $w$ limit (for square--well potentials) we find
\begin{eqnarray}
Q_{YZ} &=& {z_{YZ}\over  \Lambda_Y^2  \Lambda_Z^2} {\Omega_{YZ}  v_0 \over \sin(\gamma_Y+\gamma_Z)}
\label{qxyteth}
\end{eqnarray}
where $\gamma_Y$ and $\gamma_Z$ are the angles between the rods and the vector linking the tethering points (${\bf d}$ in Fig.\ \ref{figAppA}), and $\Omega_{YZ}$ as defined in Fig.\ \ref{FigSys}. Finally using Eqs.\ (\ref{qxyteth}) and (\ref{DG0}) we can compute $\Delta G_{YZ}$
\begin{eqnarray} 
\Delta G_{YZ} &=& \ln \Big({Q_{YZ}\over Q_{Y} Q_{Z}} \Big) = -\Delta G^0_{YZ} +\ln \Big({1\over \rho_0} {1\over \sin(\gamma_Y+\gamma_Z)} {\Omega_{YZ}\over \Omega_Y \Omega_Z} \Big)\, .
\label{result}
\end{eqnarray}
It is important to note that the $1/\sin(\gamma_Y +\gamma_Z)$ term which appears in Eq.\ (\ref{result}) is specific to the case in which the sticky ends are modelled as point particles, but that it has no physical meaning. For the purposes of the present work, different prefactors give tiny differences which do not affect any of the presented results (compare black symbols in Fig.\ \ref{FigSequences}).

 The hybridisation free energies of the sticky ends in solution $\Delta G^0_{XY}$  ($\Delta G_{0\al}$ and $\Delta G_{0\be}$ in the main text) have been computed as reported in \cite{DINAMeltRef} and implemented on the DINAMelt server \cite{DINAMeltURL} using salt concentrations equal to $[Na^+]=60\,$mM and $[Mg^{++}]=0\, $mM.

\section{Linkage swapping MC moves}\label{AppMC}
\begin{figure}
\vspace{1.5cm}
\includegraphics[angle=0,scale=0.4]{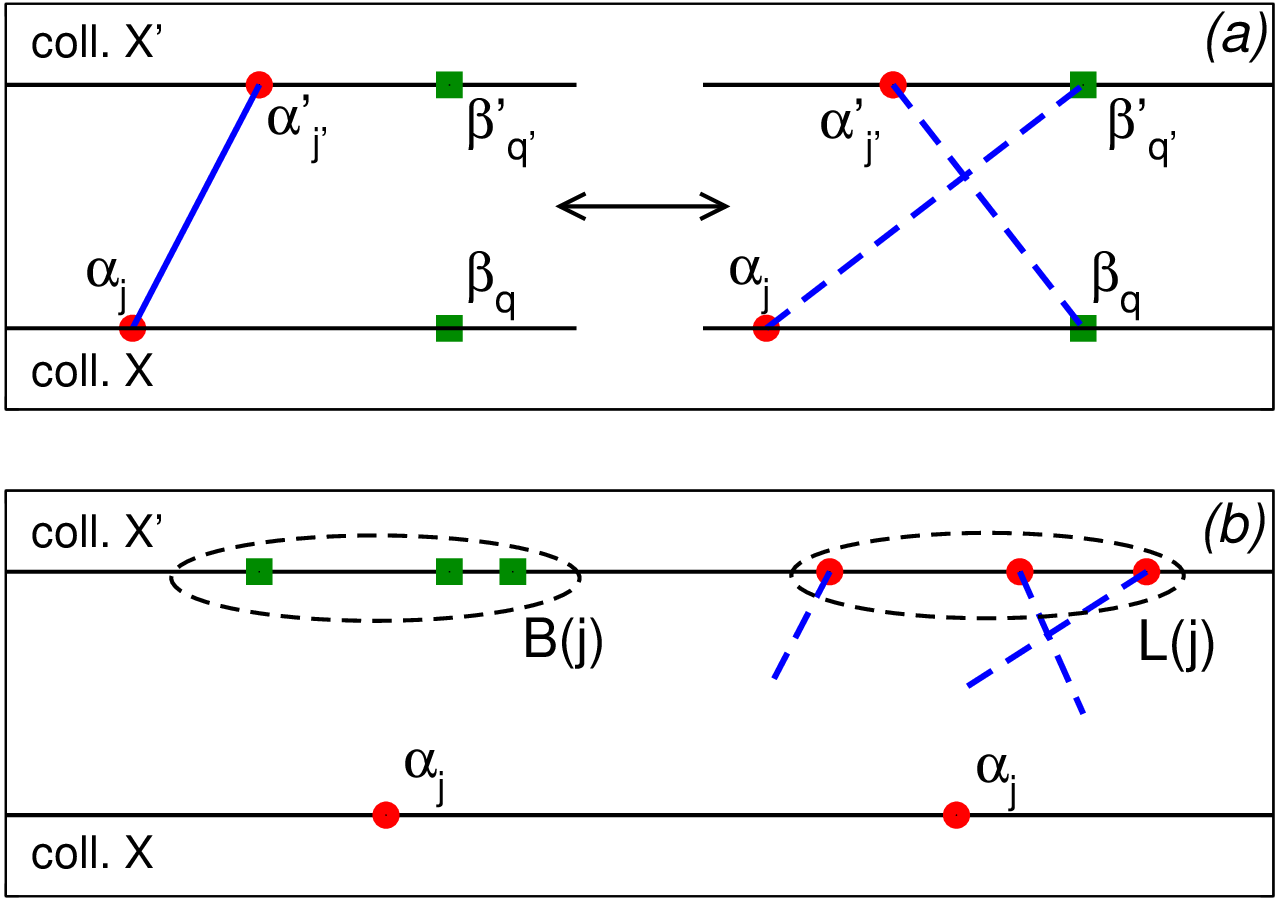} \\
\vspace{1.5cm}
\caption{(a) Implementation of a MC move between competing linkages: $\al$--$\al'$ (left) and $\al$--$\be'$+$\al'$--$\be$ (right). (b) $B(j)$ is defined to be the number of unhybridised $\be'$ strands that could be reached by $\al_j$. $L(j)$ is the number of $\al'_{j'}$ which are hybridised by a $\be$ and that could be reached by $\al_j$. Similar definitions hold for $L'(j')$ and $B'(j')$.}\label{MCfig}
\end{figure}

Here, we discuss a Monte Carlo scheme which switches between a strong linkage $\al$--$\al'$ and two weak ones $\al$--$\be'$+$\al'$--$\be$ (see Fig.\ \ref{MCfig}). In the $\al$--$\al'\to\al$--$\be'$+$\al'$--$\be$ move, we randomly select an $\al_j$--$\al'_{j'}$ linkage and two unhybridised $\be$ strands ($\be'_{q'}$, $\be_q$) from the $B(j)$ $B'(j')$ that could be connected to $\al_j$ and $\al'_{j'}$ (Fig.\ \ref{MCfig}). For the reverse move $\al$--$\be'$+$\al'$--$\be\to\al$--$\al'$ we use one of two similar schemes: A) We randomly choose two hybridised couples $\al_j$--$\be'_{q'}+\al'_{j'}$--$\be_q$ within the possible $N_\mathrm{couple}$  with $\al_j$ and $\al'_{j'}$ neighbours or B) we randomly choose an $\al_j$--$\be'_{q'}$ ($\al'_{j'}$--$\be_q$) hybridised pair from the $n_{\be,1}$ ($n_{\be',2}$) set ($n_\be=n_{\be,1}+n_{\be',2}$) and, subsequently, an $\al'_{j'}$--$\be_q$ ($\al_j$--$\be'_{q'}$) from the $L(j)$ ($L'(j')$) set with $\al_j$ and $\al'_{j'}$ that could be connected. In the two cases the acceptance rules are given by:
\begin{eqnarray}
acc_A(\al_j\al'_{j'}\to\al_j\be'_{q'}+\al'_{j'}\be_q) &=& \mathrm{min}[1,{n_\al B(j)B'(j') e^{-\beta[\Delta G(\al_j\be'_{q'})+\Delta G(\al'_{j'}\be_q)-\Delta G(\al_j\al'_{j'})]}\over N_\mathrm{couple}+L(j)+L'(j')+1}] \nonumber \\
acc_A(\al_j\be'_{q'}+\al'_{j'}\be_q\to \al_j\al'_{j'}) &=& \mathrm{min}[1,{N_\mathrm{couple} e^{\beta[\Delta G(\al_j\be'_{q'})+\Delta G(\al'_{j'}\be_q)-\Delta G(\al_j\al'_{j'})]} \over (n_\al+1) [B(j)+1][B'(j')+1] }  ] 
 \\
acc_B(\al_j\al'_{j'}\to\al_j\be'_{q'}+\al'_{j'}\be_q) &=& \mathrm{min}[1,{n_\al B(j)B'(j') e^{-\beta[\Delta G(\al_j\be'_{q'})+\Delta G(\al'_{j'}\be_q)-\Delta G(\al_j\al'_{j'})]} \over (n_{\be,1}+1)[L(j)+1]} ]
\nonumber \\
 acc_B(\al_j\be'_{q'}+\al'_{j'}\be_q\to \al_j\al'_{j'}) &=& \mathrm{min}[1,{n_{\be,1}L(j)  e^{\beta[\Delta G(\al_j\be'_{q'})+\Delta G(\al'_{j'}\be_q)-\Delta G(\al_j\al'_{j'})]} \over (n_\al+1) [B(j)+1][B'(j')+1] } ] 
\end{eqnarray}
In algorithm $B$ we randomly decide to extract first an $\al$ or $\al'$ hybridised strand.

Alternatively, we can bias the algorithm to choose couples with low configurational entropy cost. In the $\al$--$\al'\to\al$--$\be'$+$\al'$--$\be$ move, we randomly choose an $\al$--$\al'$ linkage, while $\be'_{q'}$ is selected with probability $p_{q'}$ 
\begin{eqnarray}
p_{q'} = {\exp[-\beta \Delta S^{(\mathrm{cnf})}(\al_j\be'_{q'}) ]  \over W_B(j)} &\quad \quad &  W_B(j) = \sum_{\be'_{q'}\in B(j)} \exp[-\beta \Delta S^{(\mathrm{cnf})}(\al_j\be'_{q'}) ] \quad .
\label{EqBias0}
\end{eqnarray}
Similarly, $\be_q$ is taken with probability $p_q$, where $p_q$ and $W'_B(j')$ are defined as above (\ref{EqBias0}). In the reverse move, like in algorithm $B$, we first randomly select an $\al_j\be'_{q'}$ ($\al'_{j'}\be_q$) linkage and then one between $\al'_{j'}\be_q$ ( $\al_j\be'_{q'}$)  with probability $p_{j'}$ ($p_j$), and:
\begin{eqnarray}
p_{j'} = {\exp[-\beta \Delta S^{(\mathrm{cnf})}(\al_j\al'_{j'}) ]  \over W_L(j)} &\quad \quad &  W_L(j) = \sum_{\al'_{j'}\in L(j)} \exp[-\beta \Delta S^{(\mathrm{cnf})}(\al_j\al'_{j'}) ]
\end{eqnarray}
(similar definitions follow for $p_j$ and $W'_L(j')$). Acceptance rules are then given by
\begin{eqnarray}
acc_\mathrm{bias}(\al_j\al'_{j'}\to\al_j\be'_{q'}+\al'_{j'}\be_q) &=& \mathrm{min}[1,{n_\al W_B(j) W'_{B}(j') e^{-\beta[2 \Delta G_{0\be} -\Delta G_{0\al}]} \over (n_{\be,1}+1)\overline W_L(j)} ]
\nonumber \\
 acc_\mathrm{bias}(\al_j\be'_{q'}+\al'_{j'}\be_q\to \al_j\al'_{j'}) &=& \mathrm{min}[1,{n_{\be,1} W_L(j)  e^{\beta[2 \Delta G_{0\be} -\Delta G_{0\al}]} \over (n_\al+1) \overline W_B(j)  \overline W'_{B}(j') } ] \quad,
\end{eqnarray}
where we have defined
\begin{eqnarray}
\overline W_{B}(j) = W_B(j) + \exp[-\beta \Delta S^{(\mathrm{cnf})}(\al_j\be'_{q'})]
&\qquad &
\overline W_{L}(j) = W_L(j) +\exp[-\beta \Delta S^{(\mathrm{cnf})}(\al_j\al'_{j'})]
\nonumber\\
\end{eqnarray}
(similarly for $\overline W'_{B}(j')$ and $\overline W'_{L}(j')$).

\end{document}